\documentclass{article}
\usepackage{graphicx}
\usepackage{epsfig}
\usepackage{rotating}
\usepackage{xspace}
\usepackage{lineno}
\usepackage{hyperref}
\usepackage{doi}

\newcommand{\fb}{fb$^{-1}$\xspace}

\newcommand{\ppbar}{${\rm p\bar{p}}$\xspace}
\newcommand{\ttbar}{${\rm t\bar{t}}$\xspace}
\newcommand{\ttH}{${\rm t\bar{t}H}$\xspace}
\newcommand{\tHq}{\ensuremath{\rm tHq}\xspace}

\newcommand{\pt}{$p_T$\xspace}
\newcommand{\mtop}{\ensuremath{m_{\rm t}}\xspace}
\newcommand{\thetaL}{$\theta^*_{\ell}$\xspace}
\newcommand{\costheta}{$\cos\theta^*_{\ell}$\xspace}
\newcommand{\costhetaW}{$\cos\theta^*_{\rm W}$\xspace}
\newcommand{\thetaW}{$\theta^*_{\rm W}$\xspace}
\newcommand{\phiW}{$\phi^*_{\rm W}$\xspace}
\newcommand{\etalj}{$\eta_{j'}$\xspace}
\newcommand{\vtb}{\ensuremath{|V_{\rm tb}|}\xspace}
\newcommand{\vts}{\ensuremath{|V_{\rm ts}|}\xspace}
\newcommand{\vtd}{\ensuremath{|V_{\rm td}|}\xspace}
\newcommand{\vtq}{\ensuremath{|V_{\rm tq}|}\xspace}
\newcommand{\vtbprime}{\ensuremath{|V_{\rm tb\prime}|}\xspace}
\newcommand{\MET}{\ensuremath{{E\!\!\!/}_{T}}\xspace}
\newcommand{\pol}{\ensuremath{P^{(S)}_{\rm t}}\xspace}
\newcommand{\Rb}{\ensuremath{R_{\rm b}}\xspace}
\newcommand{\RbDefOne}{\ensuremath{\frac{BR({\rm t\to Wb})}{BR({\rm t\to Wq})}}\xspace}
\newcommand{\RbDefTwo}{\ensuremath{\frac{\vtb^2}{\vtd^2+\vts^2+\vtb^2}}\xspace}
\newcommand{\Rt}{\ensuremath{R_{\rm t}}\xspace}
\newcommand{\RtDef}{\ensuremath{\sigma_{\rm t}/\sigma_{\rm \bar t}}\xspace}
\newcommand{\tbW}{${\rm tbW}$\xspace}
\newcommand{\utgamma}{${\rm ut\gamma}$\xspace}
\newcommand{\ctgamma}{${\rm ct\gamma}$\xspace}

\newcommand{\xB}{$x_B$\xspace}
\newcommand{\sigmat}{\ensuremath{\sigma^{\rm t-channel}_{\rm t/\bar t}}\xspace}
\newcommand{\de}{\ensuremath{\mathrm{d}}\xspace}
\newcommand{\mT}{\ensuremath{m_T^{\rm W}}\xspace}
\newcommand{\yt}{\ensuremath{y_{\rm t}}\xspace}
\newcommand{\BRHgg}{\ensuremath{{\rm BR(H\to\gamma\gamma)}}\xspace}

\newcommand{\somespace}{\vspace{3 mm}}

\title{Single top quark production at the LHC}

\author{Andrea~Giammanco\\
{\small Centre for Cosmology, Particle Physics and Phenomenology,} \\
{\small Universit\'e catholique de Louvain, B-1348 Louvain-la-Neuve, Belgium} \\
{\small {\bf andrea.giammanco@uclouvain.be}}
}

\begin{document}

\maketitle

\begin{abstract}
This paper is an experimental review of the study of processes with a single top quark at the LHC.
 The pioneering times are over, and this is now a sector of ``precision physics'' at colliders.
 Angular distributions of the decay products of singly-produced top quarks are unique tests of the electroweak interaction. Searches for rare final states of the form ${\rm t}+X$ (where $X={\rm \gamma, Z, H}$) are very sensitive to new physics, and will enter with Run II in a very interesting zone of the parameter space of some theories. The relative sign of the Yukawa coupling of the top quark with respect to the Higgs coupling to gauge bosons will be conclusively measured very soon in the \tHq final state. 
\end{abstract}

\section{Introduction}

The top quark is the heaviest elementary particle discovered so far, and in many ways it is a very uncommon quark. The fact that its electroweak decay is faster than the hadronisation timescale  implies that the top quark exists only as a free quark, so that the effects from new physics could show up very clearly by comparing measurements with the precise standard model (SM) predictions.
 While the pair-production process (\ttbar) was discovered more than twenty years ago~\cite{Abe:1995hr,Abachi:1995iq} and entered several years ago the domain of ``precision physics'', single top-quark production has been observed at Tevatron just two years before its shut-down~\cite{Abazov:2009ii,Aaltonen:2009jj} and precision has generally been relatively modest until recently.%, mostly because of the difficulty in the rejection and modeling of backgrounds from other SM processes.

\begin{figure}[t]
  \epsfig{file=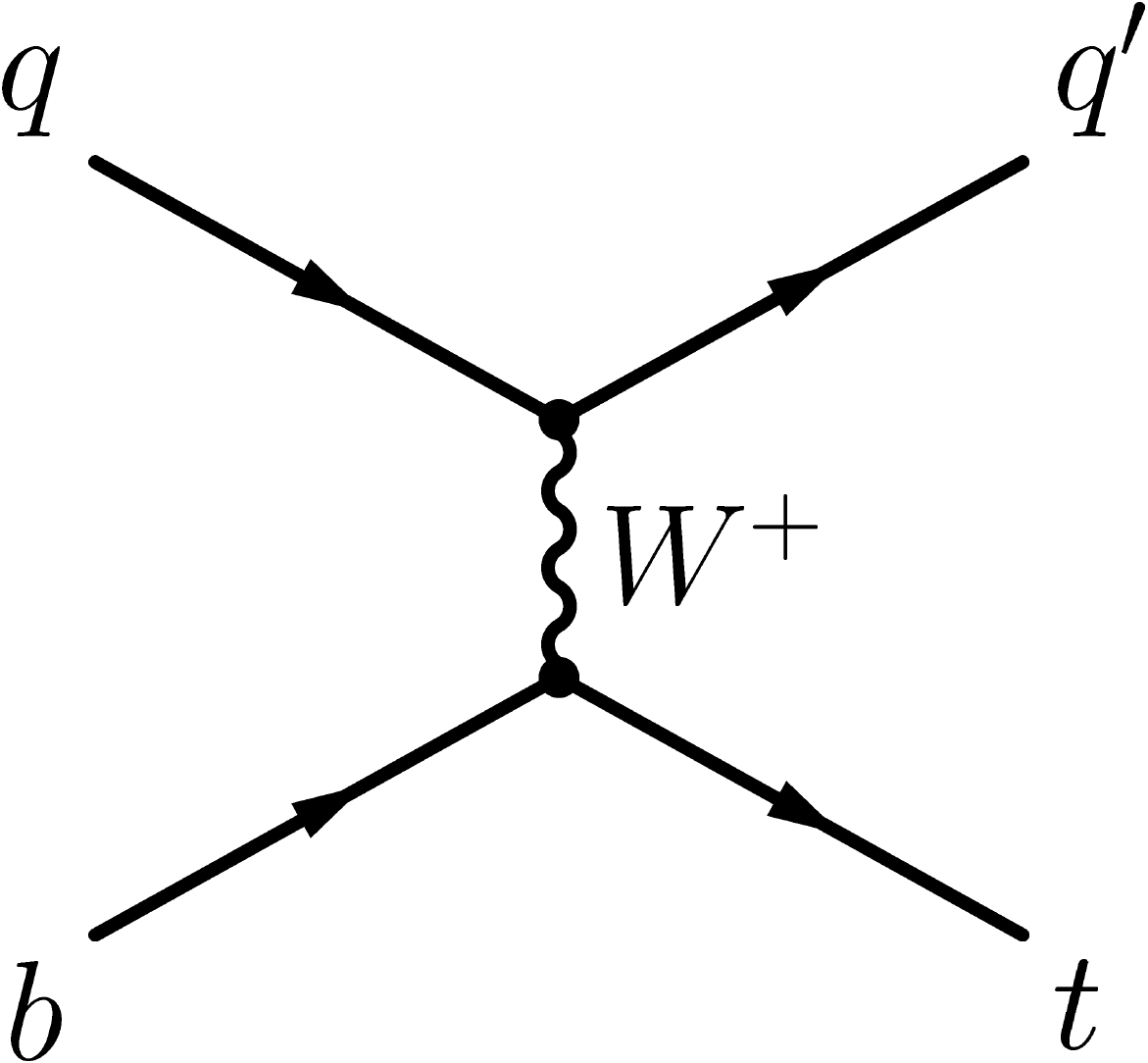,width=0.3\textwidth}
  \hfill
  \epsfig{file=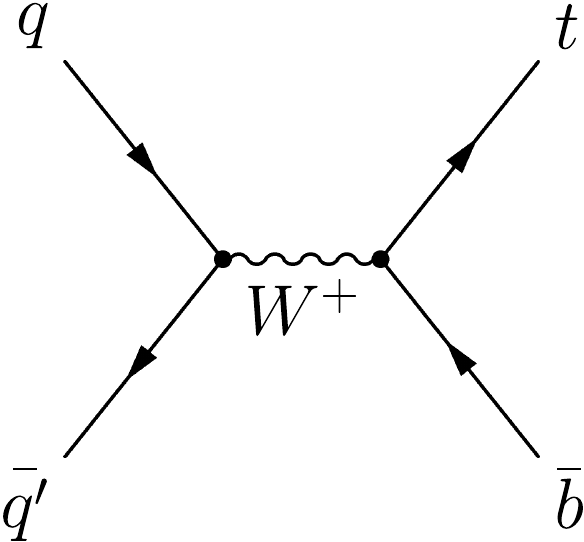,width=0.3\textwidth}
  \hfill
  \epsfig{file=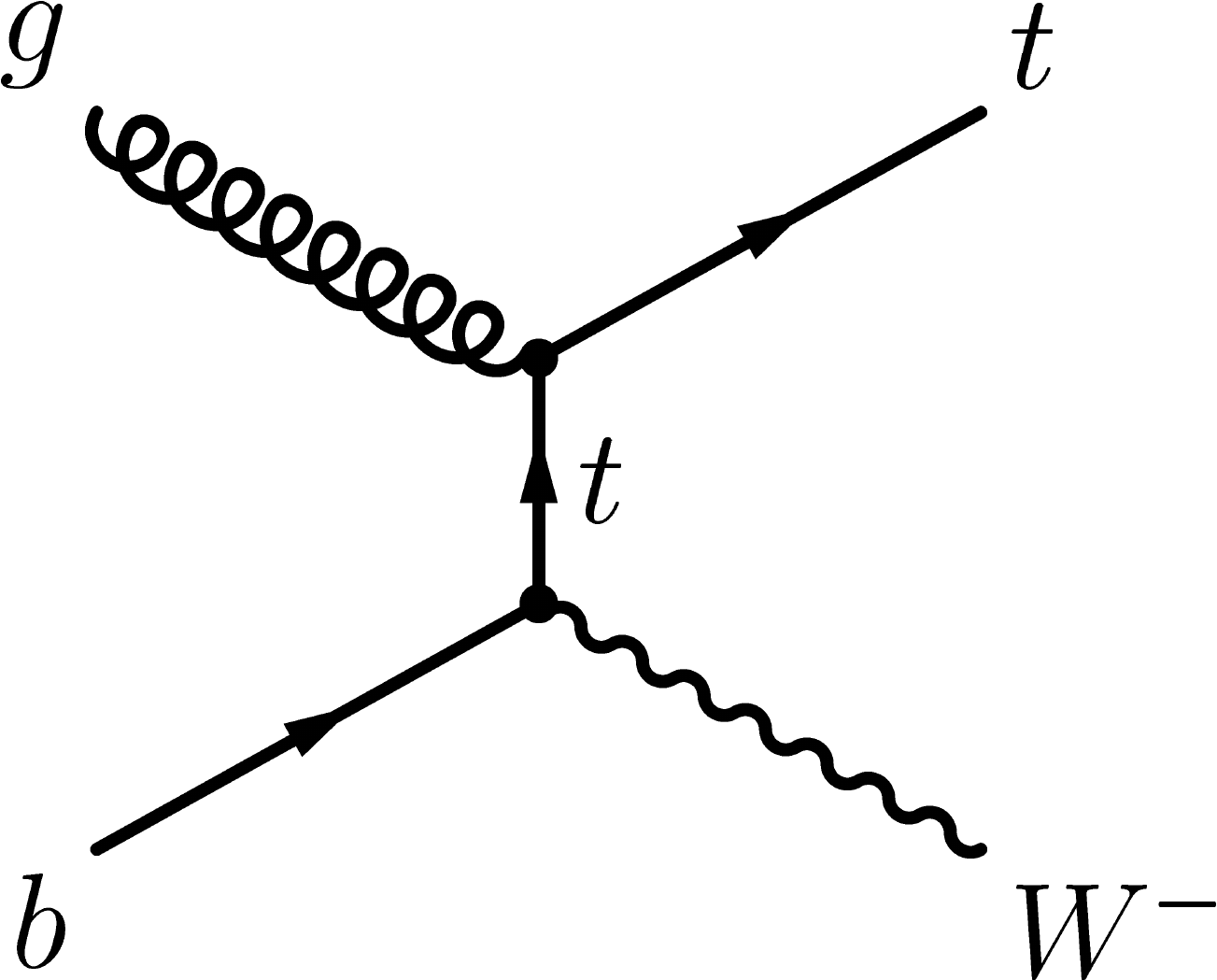,width=0.3\textwidth}
  \caption{Representative diagrams for the ``canonical channels'' of electroweak single top quark production in the standard model: t-channel (left), s-channel (centre), and W-associated production or tW (right).}
  \label{fig:FG}
\end{figure}

During the first long run of LHC (Run I), the ATLAS~\cite{Aad:2008zzm} and CMS~\cite{Chatrchyan:2008aa} experiments recorded more than 5 \fb\ of pp collisions at 7~TeV and 20 \fb\ at 8~TeV, each. 
The ``re-discovery'' of single top quark production with the data recorded at 7~TeV in 2010~\cite{Chatrchyan:2011vp} and in early 2011~\cite{Aad:2012ux} was hailed as a major milestone for these multi-purpose experiments, as it was an implicit validation of their entire data-processing chain and gave thus confidence in their studies of similarly complex final states.

%%%%%%% REDUCED UPON REFEREES CRITICISM:
Different beyond-SM models predict different effects in several production channels~\cite{Tait:2000sh}, and this motivates the study of all of them, in conjunction with \ttbar\ properties, to disentangle their effects.
 The three most abundant and most studied single top quark processes at the LHC, here referred to as the ``canonical channels'', are illustrated at Born level in Fig.~\ref{fig:FG}.
Some of these new-physics effects in t-channel and tW production might be mimicked by inaccuracies in the gluon or b-quark parton distribution functions (PDF) at large \xB~\footnote{With this symbol, this paper indicates ``Bjorken $x$'', i.e. the fraction of the incoming proton's total momentum involved in the parton-level scattering.} and it is therefore necessary to rule out this possibility by additional dedicated inputs. 
 Precise measurements of all three canonical production modes may have a deep impact  on PDF constraints, with the three channels being complementary to each other and also to \ttbar\ production.
% For example, t-channel and tW cross sections are sensitive to the b-quark PDF and anticorrelated with the W/Z cross section, while the s channel (essentially a Drell-Yan process) is insensitive to the b-quark PDF and can therefore act as a control process, and it is correlated with the W/Z cross section~\cite{Guffanti:2010yu}.
 Moreover, the integrated or differential charge asymmetry in t-channel production will provide a very powerful input for constraining PDFs, similar to the W-boson production case, in a region of \xB very relevant for several searches.

This paper elaborates on the current experimental knowledge of the properties of singly-produced top quarks after LHC Run I.
 Inclusive and differential cross sections of the three canonical processes are presented in Section~\ref{sec:xsec}.
 Section~\ref{sec:vtb} describes the extraction of \vtb\ from the single-top cross section under some assumptions, which are critically discussed.
Section~\ref{sec:anomalous} is devoted to the constraints on anomalous top-quark couplings that can be set by signatures with a single top quark. %; sub-section~\ref{sec:polarization} is specifically devoted to properties that are accessible through angular observables in t-channel production.
 Section~\ref{sec:tHq} is entirely devoted to Higgs-associated production.
 Finally, Section~\ref{sec:outlook} mentions the prospects for the analysis of Run II data.

Other related topics, like searches for ${\rm t\bar b}$ resonances~\cite{Aad:2012ej,Aad:2014xra,Aad:2014xea,Chatrchyan:2014koa,Khachatryan:2015edz}, singly produced new heavy quarks~\cite{Aad:2013rna,Aad:2014efa,Aad:2015voa,Khachatryan:2015mta}, single top quarks in association with invisible new particles~\cite{Aad:2014wza,Khachatryan:2014uma}, are out of the scope of this paper.

\section{Cross sections of the canonical channels}
\label{sec:xsec}

\begin{figure}[!h]
 \begin{center}
  \includegraphics[width=0.95\textwidth]{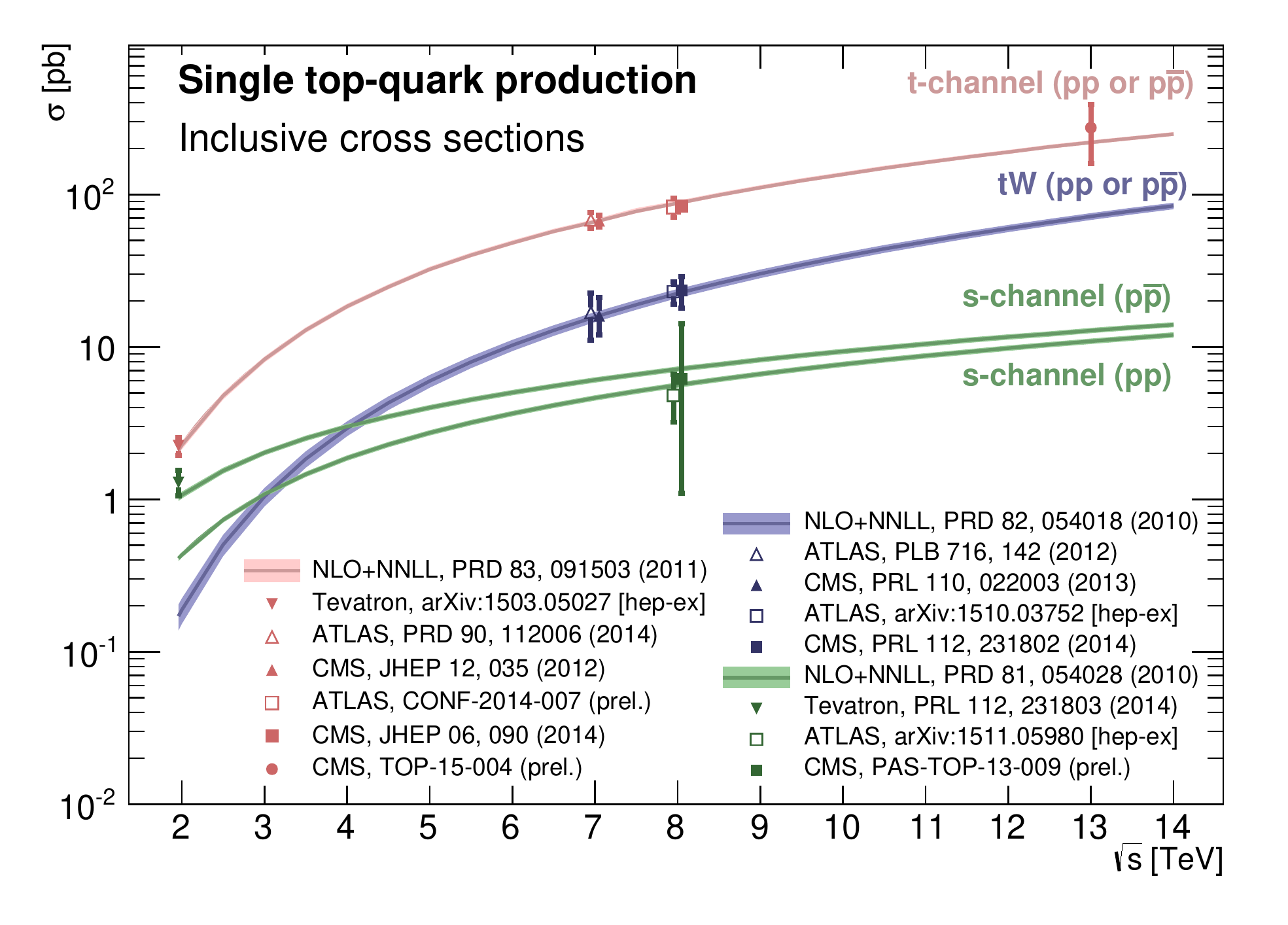}
  \caption{\label{fig:sqrts}{} Summary of Tevatron and LHC measurements of the inclusive single top-quark production cross sections in t-channel, s-channel and tW production. The measurements, which include some preliminary results, are compared to theoretical calculations based on NLO QCD complemented with NNLL resummation. The full theory curves as functions of the centre of mass energy are calculated as in Refs.~\cite{Kidonakis:2011wy,Kidonakis:2010ux,Kidonakis:2010tc}.}
 \end{center}
\end{figure}

Figure~\ref{fig:sqrts} shows the state of the art for the experimental measurements of the inclusive cross section of the three main single top-quark production mechanisms (Fig.~\ref{fig:FG}) at Tevatron and LHC centre-of-mass energies.

\somespace
The electroweak t-channel mode of production (Fig.~\ref{fig:FG}, left) being the most abundant is also the most studied. This is the only single top-quark process whose cross section has been measured at four centre-of-mass energies so far.
%~\footnote{In fact, four if one considers unpublished results that have been made public as preliminary results for conferences, see Fig.~\ref{fig:sqrts}.}.
 Predictions have been calculated at next-to-next-to-leading order (NNLO) in quantum chromodynamics (QCD)~\cite{Brucherseifer:2014ama} and at next-to-leading order (NLO) with next-to-next-to-leading logarithm resummation (NNLL)~\cite{Kidonakis:2011wy}. Automatic calculations as a function of various parameters can be performed with the {\sc HATHOR v2.1} program at NLO~\cite{Aliev:2010zk,Kant:2014oha}.
 At the Tevatron, the inclusive cross section has been measured at 1.96~TeV~\cite{Aaltonen:2015cra} by CDF and D0 in $\rm p\bar p$ collisions~\cite{Aaltonen:2015cra}. 

At the LHC, inclusive t-channel cross sections have been measured at 7~TeV~\cite{Aad:2014fwa,Chatrchyan:2012ep} and 8 TeV~\cite{Khachatryan:2014iya} by ATLAS and CMS. All these analyses enhance the t-channel signal by selecting events with one isolated electron or muon, significant missing transverse energy (\MET) and/or large transverse invariant mass (\mT) of the lepton plus \MET system~\footnote{Defined as $\mT = \sqrt{ \left(p_T^{l} + \MET \right)^2 - \left( p_x^{l} + \MET{}_{,x} \right)^2 - \left( p_y^{l} + \MET{}_{,y} \right)^2 }$. Here and anywhere in this article, symbol $l$ is used to refer to a charged lepton, $p_x$ and $p_y$ indicate momentum components along the $x$ and $y$ axis chosen as orthogonal directions to the beam axis, and $p_T \equiv \sqrt{p_x^2+p_y^2}$ (transverse momentum).}, and two or three jets. Exactly one of the jets is required to pass a tight threshold on the likelihood of originating from a b
quark (b-tagged jet) and is interpreted as coming from the decay of the top quark, and the other (failing the same threshold) as originating from the spectator quark that recoils again the top. Main backgrounds to this final state are \ttbar\ and W boson produced in association with jets (W+jets). Orthogonal control regions with different multiplicities of jets and/or b-tagged jets are used to measure these backgrounds {\it in situ} or to validate the Monte Carlo models used for their predictions,  or to constrain the main experimental systematics (e.g., b-tagging efficiency).
 QCD multi-jet events constitute a non-negligible background. Given the uncertainties in its modelling~\footnote{Processes in this category have a very low event survival probability given the tight thresholds of these selections but also a very large cross section. Therefore, huge Monte Carlo samples would be needed to sufficiently populate the tails of their distributions such to give fairly precise predictions in our signal regions, often beyond the current computing resources of the experimental collaborations. Moreover, a satisfactory prediction of these extreme tails would demand an exceptional level of accuracy of the full simulation chain.} it is necessary to predict the size and properties of this process by data themselves. A reliable model of this background is usually extracted by events that fail the isolation requirement or other elements of the charged-lepton selection, while fulfilling all other selection criteria.
 The extraction of the signal cross section is performed by both collaborations by profile likelihood fits. The fit variable is a multivariate discriminant in the case of ATLAS~\cite{Aad:2014fwa} and of some of the CMS analyses~\cite{Chatrchyan:2011vp,Chatrchyan:2012ep}. CMS also demonstrated the feasibility of entirely relying on a simple kinematic observable, \etalj, the pseudorapidity of the jet failing b tagging~\cite{Chatrchyan:2012ep,Khachatryan:2014iya}. %This is referred to as ``\etalj analysis'' in the following.
 The motivation for this different strategy is the wish to remove this implicit SM assumption, as it relies on a kinematic property of the signal events that is not significantly affected by the properties of the \tbW coupling~\footnote{The \etalj distribution is mostly shaped by the mass of the heavy system that the light quark is recoiling against, and by the light-quark PDF in the proton, which is relatively well known in the relevant \xB region.}.

Differential cross sections of t-channel production as a function of top-quark \pt\ and pseudorapidity have been measured by ATLAS at 7~TeV~\cite{Aad:2014fwa},
% and CMS at 8~TeV~\cite{CMS-PAS-TOP-14-004},
 showing a good agreement with the Standard Model predictions of various MC generators.

\begin{figure}[t]
  \epsfig{file=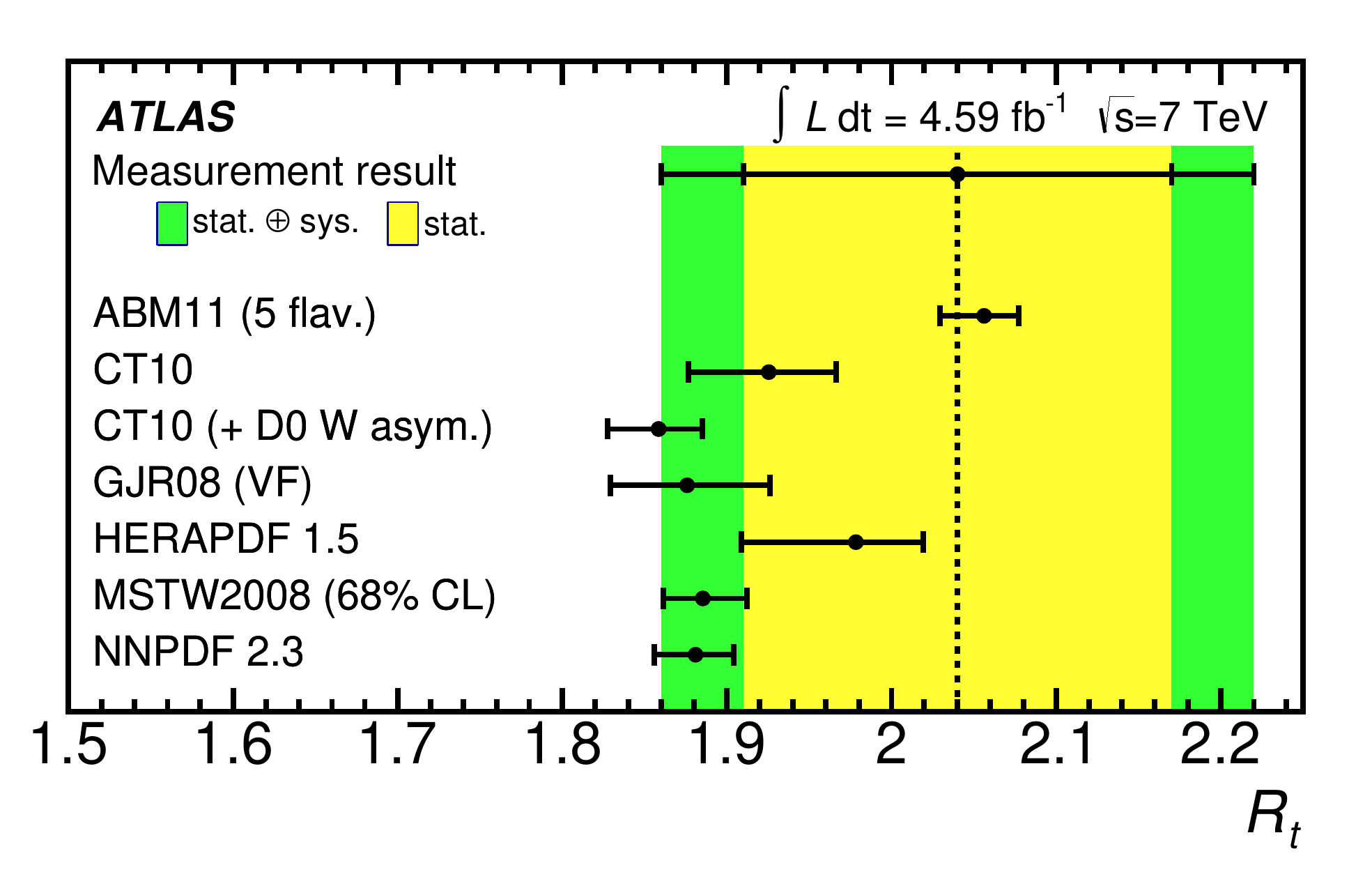,width=0.55\textwidth}
  \epsfig{file=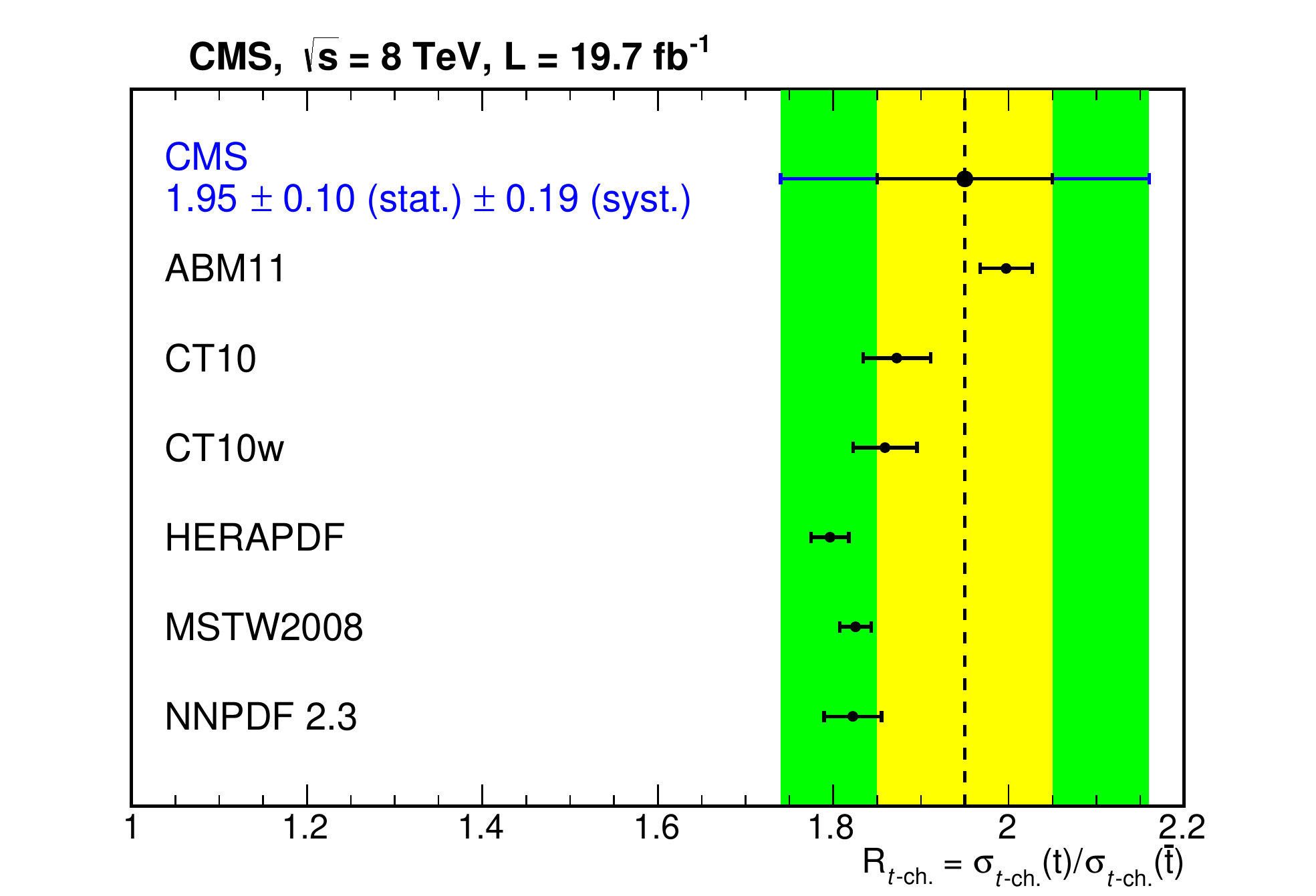,width=0.45\textwidth}
  \caption{Values of \Rt measured by ATLAS at 7~TeV~\cite{Aad:2014fwa} (left) and CMS at 8~TeV~\cite{Khachatryan:2014iya} (right), compared to the values calculated for different NLO PDF sets~\cite{ABM11,CT10,GJR08,HERAPDF,MSTW2008NLO,NNPDF}. The error on the calculated values contains the uncertainty on the renormalisation and factorisation scales. 
 The combined statistical and systematic uncertainties of the measurement are shown by the green band added to the yellow band, which shows the statistical errors only.}
  \label{fig:tchan_ratio}
\end{figure}

%%% ratio
 A feature of SM single-top production at the LHC is the difference in production rate (integrated charge asymmetry, $\Rt\equiv \RtDef$) between top and anti-top production in t and s channels.
 The measured values of \Rt from the ATLAS collaboration at 7~TeV~\cite{Aad:2014fwa} and the CMS collaboration at 8~TeV~\cite{Khachatryan:2014iya} show qualitatively similar trends and are already disfavouring some PDF sets, as illustrated in Fig.~\ref{fig:tchan_ratio}. With more data, differential distributions of \Rt as a function of rapidity and transverse momentum of the top quark will provide significant additional discriminating power.
 It would be interesting to verify the SM prediction for single top-quark production in \ppbar collisions, $\Rt = 0$, but the Tevatron data set is insufficient for a measurement of \Rt with interesting precision.

Another useful input for constraining PDFs is the measurement of the ratio of single-top cross sections between 7 and 8 TeV, as done by the CMS collaboration in the t-channel case by taking the ratio of results of the \etalj-based analysis at the two centre-of-mass energies~\cite{Khachatryan:2014iya}.

\somespace
 The first evidence of tW production has been reported by the ATLAS and CMS collaborations using 7~TeV data~\cite{Aad:2012xca,Chatrchyan:2012zca}. The conventional $5\sigma$ threshold has been crossed with 8~TeV data~\cite{Chatrchyan:2014tua,Aad:2015eto}. %~\cite{TheATLAScollaboration:2013fja,Chatrchyan:2014tua}.
% A combination of the latter measurement with a preliminary ATLAS measurement at 8~TeV has been performed by the two collaborations in the framework of the Top LHC Working Group~\cite{combo_tw} and is in agreement with a Standard Model calculation at NLO in QCD~\cite{Kidonakis:2013zqa}.
 These analyses exploit the presence of two real W bosons (the associated one, and the one from top-quark decay) by selecting events with two charged leptons (electrons or muons).
 Both collaborations perform a profile likelihood fit on a multivariate discriminant.
 Selections with one or two jets, out of which one or two b-tagged, are used to simultaneously extract the amount of signal (mostly in the final state with only one jet, passing b-tagging selection) and of the most abundant background process, \ttbar, in order to minimise the effect of its systematic uncertainties on the signal cross section measurement. 

Much theoretical work has been devoted to the issue of the quantum interference between the tW process, which is well-defined only at Born level, e.g. as in Fig.~\ref{fig:FG}(right), and \ttbar when higher-order QCD diagrams are taken into account (see, for example, Refs.~\cite{Belyaev:2000me,Campbell:2005bb,Frixione:2008yi}).
 The NLO event generators {\sc MC@NLO}~\cite{Frixione:2002ik} and {\sc POWHEG}~\cite{Frixione:2007vw}
 allow to choose between the so called ``Diagram Removal'' (DR) and ``Diagram Subtraction'' (DS) approaches~\cite{Frixione:2008yi,White:2009yt,Re:2010bp}. The DR approach consists in removing all diagrams where the associated W boson and an associated b quark form a top quark; the DS approach makes use of a subtraction term designed to locally cancel the \ttbar contributions. While the latter approach is designed such to be gauge-invariant, the former breaks gauge invariance explicitly, but this is demonstrated to have little practical effect.
 ATLAS and CMS analyses are tailored for the Born-level description of the tW process, but a specific systematic uncertainty is assigned by looking at the difference between using DR and DS signal models.

 To reduce the dependence on the theory assumptions, ATLAS also reports a cross section in a fiducial detector acceptance defined by the presence of two charged leptons and exactly one b jet at particle level~\cite{Aad:2015eto}. The definition of signal encompasses not only tW production but also \ttbar production where one of the final-state b quarks is outside of acceptance. The result is found in agreement with NLO predictions from two different matrix-element generators ({\sc POWHEG} and {\sc MC@NLO}), two different parton-shower generators ({\sc Pythia~6}~\cite{Sjostrand:2006za} and {\sc Herwig~6}~\cite{Corcella:2000bw}), the DR and DS approaches, and a variety of PDF sets. Results are also compatible with a {\sc POWHEG} prediction where tW is calculated at NLO+NNLL accuracy and \ttbar at NNLO+NNLL accuracy.

\somespace
The production of s-channel single top quarks (Fig.~\ref{fig:FG}, middle) is initiated, at Born level, by ${\rm q\bar q^\prime}$ annihilation and is therefore more abundantly produced in \ppbar than in pp collisions, as visible in Fig.~\ref{fig:sqrts}, differently from t-channel and tW production.
 Its existence has been established very recently by the combination of Tevatron measurements~\cite{CDF:2014uma} and it is one of the few ``Tevatron legacies'' that have not been surpassed in precision by the LHC experiments.
 The ATLAS collaboration recently confirmed the evidence with the 8~TeV dataset~\cite{Aad:2015upn}. %, followed by CMS~\cite{CMS-PAS-TOP-13-009}.
 These analyses select events with one isolated electron or muon, significant missing transverse energy (\MET) and/or large \mT, and two jets, both b-tagged. 
 Main backgrounds are \ttbar, W+jets, QCD multi-jet production, and the other single top-quark processes.
 Several orthogonal control regions with different multiplicities of jets and/or b-tagged jets are used to measure these backgrounds {\it in situ} or to validate the Monte Carlo models used for their predictions,  or to constrain the main experimental systematics (e.g., b-tagging efficiency).
 Due to the difficulty of the analysis problem, advanced multivariate techniques like the Matrix Element Method~\cite{Kondo:1988yd} are crucial in the study of this process.

\section{Single top quarks and \vtb}
\label{sec:vtb}

\begin{figure}[!h]
 \begin{center}
  \includegraphics[width=0.9\textwidth]{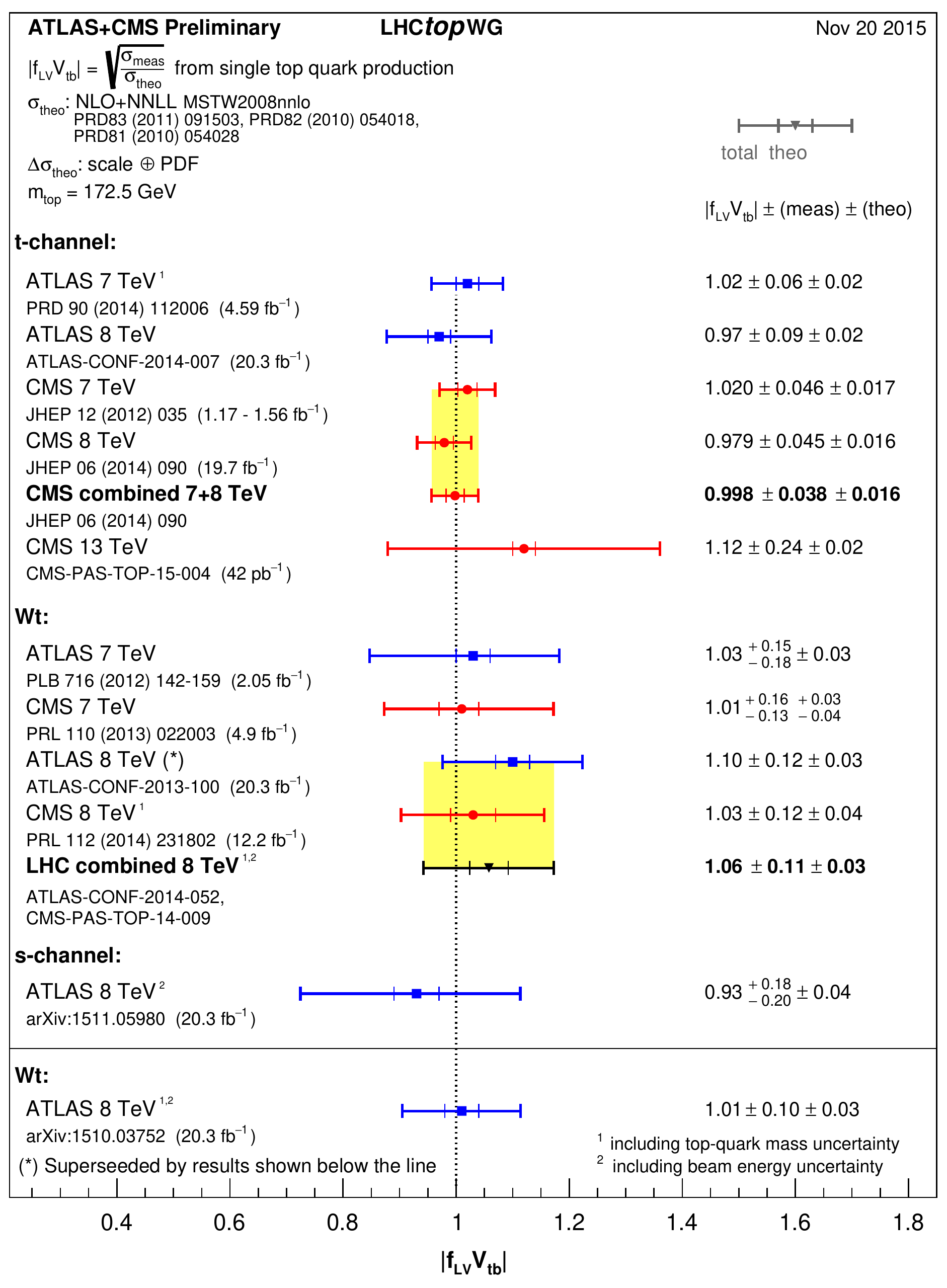}
  \caption{\label{fig:vtb}{} Summary of ATLAS and CMS extractions of \vtb from the single top-quark cross-section measurements, using NLO+NNLL theoretical predictions. From the LHC Top Working Group~\cite{toplhcwg}, including some preliminary results.}
 \end{center}
\end{figure}

Most of the single top-quark cross section measurements published so far include an inference of the \vtb\ interval corresponding to the measured cross section, under the simplifying assumption that, whatever the values, the relationships $\vtb \gg \vtd$ and $\vtb\gg \vts$ hold true, which makes the cross section of the processes in Fig.~\ref{fig:FG} directly proportional to $\vtb^2$. Also implicit is the assumption of purely left-handed \tbW coupling.
 Figure~\ref{fig:vtb} shows the \vtb\ intervals extracted by the LHC experiments under these assumptions~\cite{toplhcwg}. It is customary to also quote the 95\% confidence level interval obtained with the additional unitarity constraint $0\le \vtb\le 1$.
 This way to constrain \vtb is complementary to the measurement of
%The most precise determination from single top~\cite{Khachatryan:2014iya} has twice the uncertainty of the best one derived from 
 $\Rb\equiv \RbDefOne$ in \ttbar events~\cite{Abazov:2011zk,Aaltonen:2013luz,Aaltonen:2014yua,Khachatryan:2014nda}, where ${\rm q=d,s,b}$, which is interpreted as $\Rb = \RbDefTwo$ and can be used to infer \vtb\ directly if the additional assumption $\vtd^2+\vts^2+\vtb^2 = 1$ is imposed.
 The most precise measurement of $\Rb$, under these assumptions, yields $\vtb = 1.007\pm 0.016$~\cite{Khachatryan:2014nda}.

Some papers~\cite{Tait:2000sh,Alwall:2006bx,Lacker:2012ek} examined how single-top cross sections are modified under the hypothesis of additional quarks that mix with the known up-type quarks, and therefore of a violation of unitarity for the $3\times 3$ components of the extended ($N\times N$, with $N>3$) CKM matrix.
 While the sum $\vtd^2+\vts^2+\vtb^2+\vtbprime^2$ and, a fortiori, the sum $\vtd^2+\vts^2+\vtb^2$ is bound to be $\le 1$ also in the extended matrix, the constraints on \vtd and \vts derived from precision physics~\cite{Agashe:2014kda} do not hold when their underlying assumptions (e.g., no non-SM particles in the loops) are relaxed~\cite{Alwall:2006bx}.
 Therefore, contrary to common intuition, additional quark generations could result in an {\it enhancement}, and not only necessarily a deficit, of the single top-quark cross section in t-channel and tW productions, due to the large parton densities of d and s quarks in the proton (much larger than the b-quark density), causing an amplification of the effect of any \vtd or \vts value larger than the SM expectation, which can overcome the deficit due to a smaller \vtb.

 It is worth remarking that mixing may happen not only with sequential replicas of the known quarks (easily accommodated in the SM framework, but severely constrained by the Higgs cross section measurements~\cite{Lenz:2013iha}) but in general with any hypothetical quark-like particle with the appropriate quantum numbers. Moreover, the effective mixing matrix may be rectangular. An interesting example are vector-like quarks~\cite{Okada:2012gy,Aguilar-Saavedra:2013qpa}.

References~\cite{Alwall:2006bx,Lacker:2012ek} performed the exercise of deriving less model-dependent limits on all three \vtq matrix elements by re-examining the measurements of single top-quark cross sections and \Rb published at the time.
 This typically imposes several approximations and short-cuts, not having access to the raw data.
 A particularly tricky case for the reinterpretation are single-top analyses based on multivariate techniques, because several of the input variables are related to the kinematics of a reconstructed top quark, and the choice of the jet that is assumed to come from the decay of the top quark is made under the assumption that it is a b jet, and therefore $\vtb\approx 1$.
 It is recommended that the experimental collaborations carry out such an exercise
themselves, with minimal model dependence at every stage of the analysis, training multivariate discriminators to distinguish between different \vtb scenarios.

\section{Anomalous top-quark couplings}
\label{sec:anomalous}

The fact that the mass of the top quark is of the order of the electroweak symmetry-breaking scale (in other terms, that $|\yt|\approx 1$, where $y_t$ is the top quark Yukawa coupling) motivates several beyond-SM models that try to find a natural explanation for that, e.g. top flavour models with seesaw mechanism~\cite{He:1999vp}, top colour seesaw models~\cite{Dobrescu:1997nm}, models with vector-like quarks~\cite{Okada:2012gy}.
 These models, by assigning some special role to the top quark, typically predict larger anomalies in the top-quark sector than for other quarks.
 Complementary to direct searches for ``top-quark partners'' or other striking signatures of those models, many analyses of collider data set constraints on so called ``anomalous couplings'' and/or the parameters of effective field theories (EFT), that can be easily re-casted in terms of the fundamental parameters of large families of explicit models.

For example, the most general Lagrangian term that one can write for the \tbW coupling up to dimension-six gauge invariant operators~\cite{AguilarSaavedra:2008zc} (under the approximation $\vtb = 1$) is:
\begin{equation}
\label{eq:tbW}
\mathcal{L}_{\rm tbW} = -\frac{g}{\sqrt{2}}\bar b\left[ \gamma^{\mu}(f_L P_L + f_R P_R) + \frac{i\sigma^{\mu\nu}q_{\nu}}{M_{\rm W}} (g_L P_L + g_R P_R) \right] t W^-_{\mu} + h.c. \, ,
\end{equation}
where the parameters $f_L$ and $f_R$ denote the strength of the left- and right-handed vector-like couplings, and $g_L$ and $g_R$ denote the left- and right-handed tensor-like couplings.
 The SM predicts $f_L = 1$, $f_R = g_L = g_R = 0$.

\subsection{Flavour Changing Neutral Current signatures}

 Models that try to solve the so called ``flavour problem''~\cite{Georgi:1986ku} usually predict a large coupling of new particles to the top quark, and therefore sizable flavour-changing neutral current (FCNC) effects in the top quark sector, despite the tight constraints in the B- and K-meson sectors. These are very interesting to look for in single top-quark production, where the effect of a small u--t coupling would be enhanced by the large u-quark density~\cite{Tait:2000sh}.
 Formulations exist where the effect of new particles in quantum loops is absorbed by effective tuX couplings, where $X$ can be a gluon, a photon, a Z or a H boson (read, for example, refs.\cite{AguilarSaavedra:2008zc,Zhang:2010dr}). Based on the consideration that higher-order effects mix the effects of different couplings, inducing ambiguities in the interpretation of single signatures, a global approach is advocated in ref.~\cite{Durieux:2014xla}. However, the results reviewed in this paper make use of leading-order FCNC models.

\somespace
%The CMS collaboration has set constrains on several effective parameters in ref.~\cite{CMS-PAS-TOP-14-007}. After a model-independent selection of t-channel production events, several Bayesian neural networks (NN) combining several angular and kinematical inputs have been trained, for non-zero values of each of the anomalous parameters $f_R, g_L, g_R$ in Eq.~\ref{eq:tbW} and for the signature of FCNC production through a non-zero $ugt$ coupling. 

\begin{figure}[t]
  \centering
  \epsfig{file=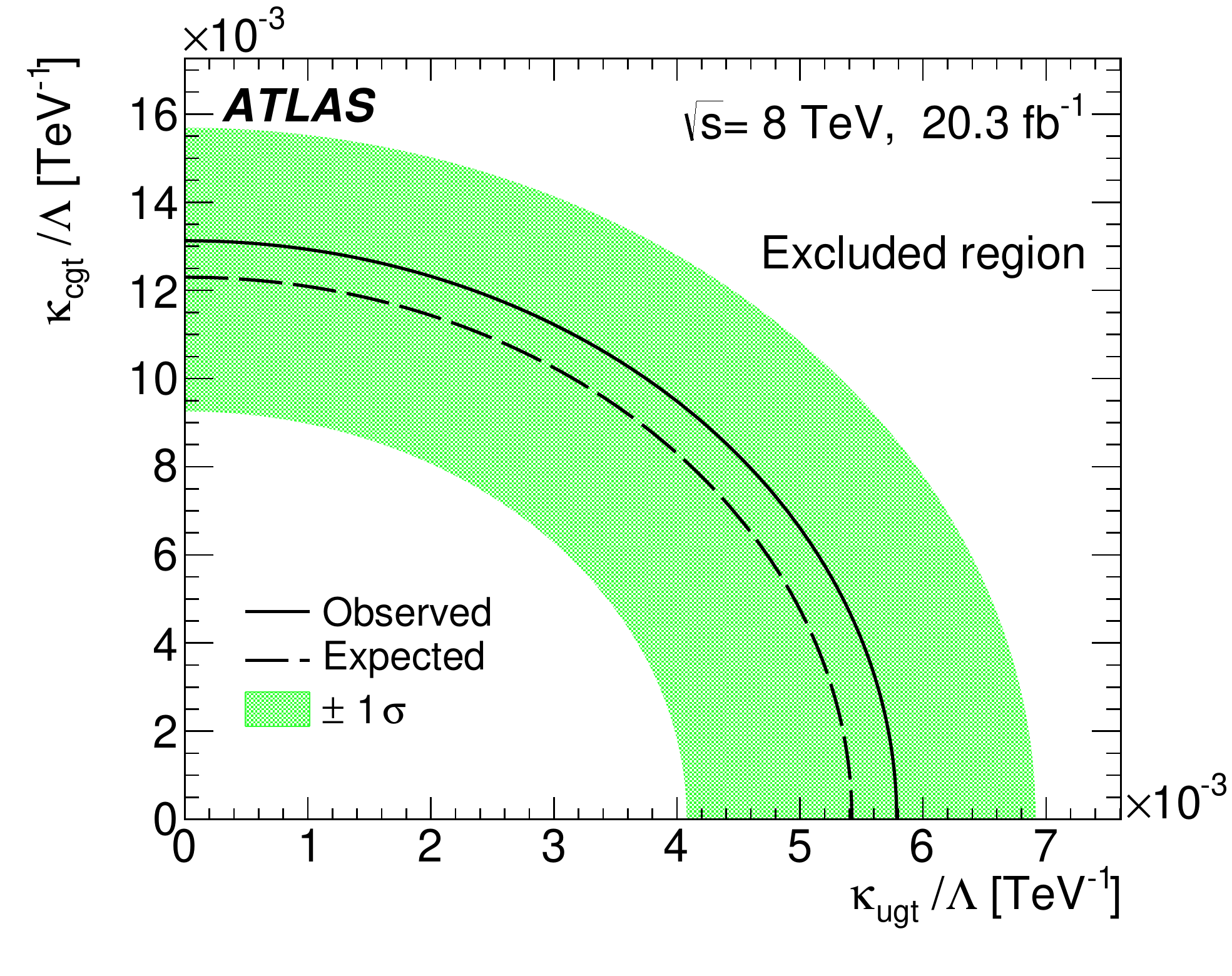,width=0.49\textwidth}
  \epsfig{file=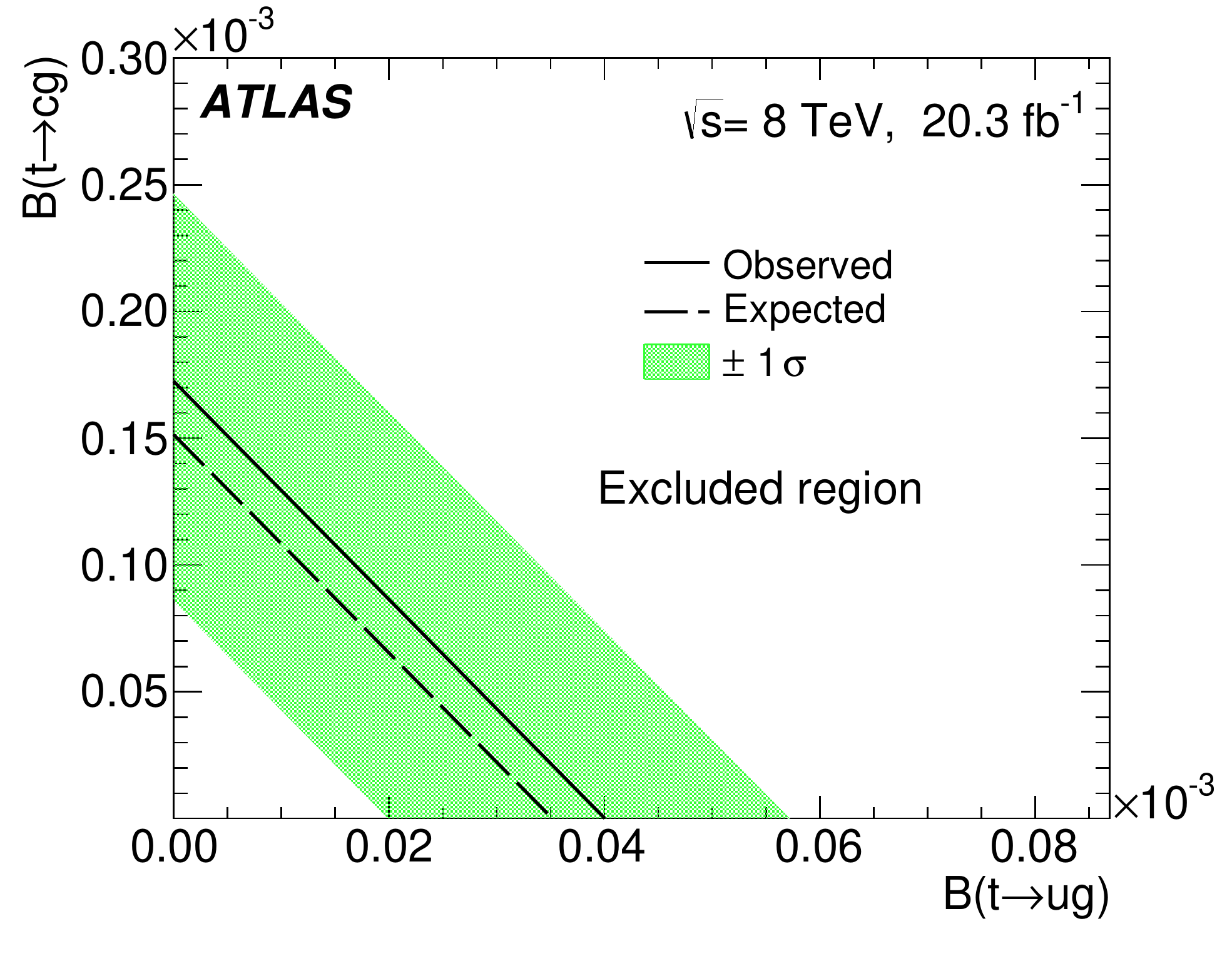,width=0.49\textwidth}
  \caption{%Left: leading order diagram for ``very single top-quark production'' by an anomalous (FCNC) coupling of the $tug$ and $tcg$ kind. 
Left: upper limit on the effective coupling constants between top quarks, gluons, and up or charm quarks, $\kappa_{\rm ugt}$ and $\kappa_{\rm cgt}$. Right: upper limit on the branching fractions ${\rm BR(t \to ug)}$ and ${\rm BR(t \to cg)}$. The shaded band shows the one standard deviation variation of the expected limit. 
 Figures from ref.~\cite{Aad:2015gea}.}
  \label{fig:very-single-top}
\end{figure}

The ATLAS collaboration searched for the exotic signature of a ``very single top quark'' (i.e., a $2\to 1$ partonic reaction producing a top quark) with the 7 and 8~TeV data sets~\cite{Aad:2012gd,Aad:2015gea}, to constrain the FCNC couplings of the utg and ctg kinds, see Fig.~\ref{fig:very-single-top}. The analysis selects events with a single charged lepton, significant \MET and a single jet, passing b-tagging identification. A Bayesian Neural Network is applied on the selected events, trained to separate FCNC signals from SM events.

%The D0 collaboration chose a different strategy, based on the SM-like signature of a single top quark in association with a light quark~\cite{d0_fcnc} (same signature as t-channel production in the SM). With a similar integrated luminosity as CDF (2.3~fb$^{-1}$, versus 2.2~fb$^{-1}$ for CDF) they obtained more stringent limits, demonstrating that the ``top+jet'' signature (mostly initiated by quark-quark collisions) is more sensitive than the ``very single top'' signature in the Tevatron conditions. Nevertheless, at the LHC energies the gluon flux is much more intense than at Tevatron in comparison to the quark flux, therefore the balance is likely in favour of the ``very single top''\footnote{A proper comparison, beyond the scope of this paper, would demand to take into account any effect susceptible to increase the jet multiplicity of the event, like initial state radiation and pileup.}.

\begin{figure}[t]
  \centering
  \epsfig{file=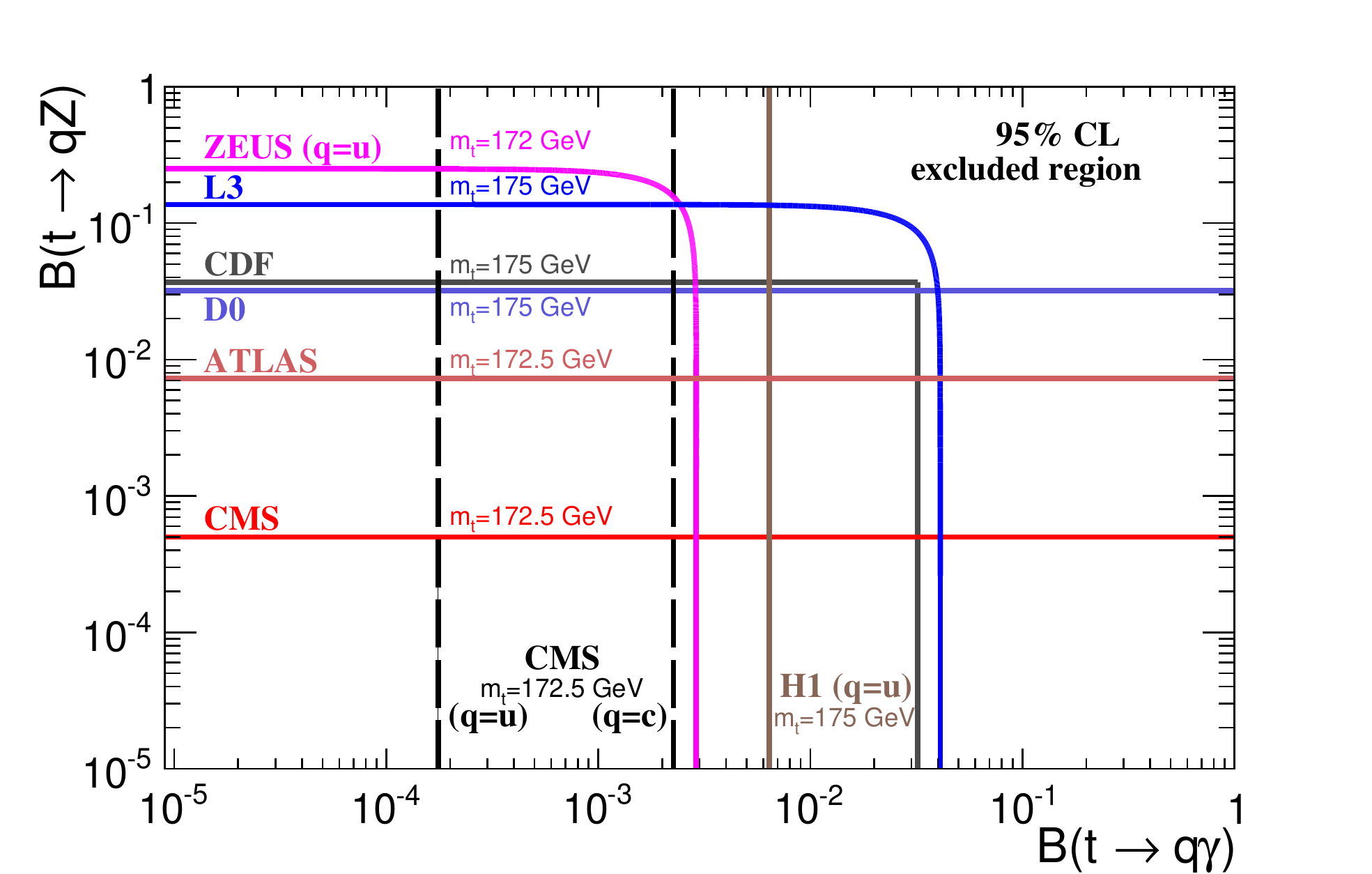,width=0.95\textwidth}
  \caption{Observed 95\% CL upper limit on the branching ratio of ${\rm t\to Zq}$ versus the branching of ${\rm t\to \gamma q}$ (${\rm q = u,c}$) as derived directly or indirectly by experiments at LEP, HERA, Tevatron and LHC: search for ${\rm e^+e^-\to \gamma^{*}/Z \to t\bar q / \bar t q}$ by L3~\cite{Achard:2002vv}, search for ${\rm eq\to et}$ by ZEUS~\cite{Abramowicz:2011tv} and H1~\cite{Aaron:2009vv}, search for ${\rm t\to Zq}$ decays in \ttbar events by D0~\cite{Abazov:2011qf}, CDF~\cite{Aaltonen:2008ac}, ATLAS~\cite{Aad:2012ij} and CMS~\cite{Chatrchyan:2013nwa}, search for ${\rm t\to \gamma q}$ decays in \ttbar events by CDF~\cite{PhysRevLett.80.2525}, search for single top quark plus photon by CMS~\cite{Khachatryan:2015att}.}
  \label{fig:tgamma}
\end{figure}

\somespace
The CMS collaboration searched for events containing a top quark and a large-\pt photon with the 8~TeV data set~\cite{Khachatryan:2015att}. The semileptonic decay of the top quark is used, and a multivariate analysis is performed to discriminate the FCNC signal from the SM backgrounds.
 The dominant W+jets and ${\rm W+\gamma+}$jets backgrounds are estimated from data.
 This statistically-limited analysis makes use of the event counts to set limits on the effective couplings of the \utgamma and \ctgamma types. For the purpose of easy comparison with measurements in \ttbar production, the result is also interpreted in terms of an equivalent branching ratio of top-quark decay into a photon and a quark, see Fig.~\ref{fig:tgamma}.

\subsection{Insights from angular correlations}
\label{sec:polarization}

As often remarked in the literature, the top quark is the only quark that can be studied ``naked''.
This comes from the very neat separation of the relevant time scales: the production time ($\approx 1/\mtop$, where $\mtop\approx 175$~GeV) is two orders of magnitude smaller than its lifetime ($1/\Gamma_{\rm t}$, where $\Gamma_{\rm t}\approx 2$~GeV) which is smaller than the hadronisation time scale ($\approx 1/\Lambda_{\rm QCD}$, where $\Lambda_{\rm QCD}\approx 0.2$~GeV) which, in turn, is an order of magnitude smaller than the spin decorrelation time ($\approx \mtop/\Lambda_{\rm QCD}^2$).
 In particular, the fact that the top quark lifetime is shorter than the QCD de-coherence timescale implies that its decay products retain memory of its helicity.
 This provides additional powerful tools in the search for new physics in single-top studies: 
 in  single top-quark production via the t channel, the standard model predicts that top quarks are produced almost fully polarised through the V--A coupling along the direction of the momentum of the quark
 that recoils against the top quark~\cite{Mahlon:1999gz,Jezabek:1994zv}, while new physics models may lead to a depolarisation in production or decay by altering the coupling structure~\cite{AguilarSaavedra:2010nx, AguilarSaavedra:2008gt,AguilarSaavedra:2008zc,Bach:2012fb}.

The general form of the angular distribution ($\theta^{*}_X$) of decay product $X$ (${\rm =W,\ell,\nu,b}$) with respect to the top-quark spin direction in the top-quark rest frame is
\begin{equation}
\frac{1}{\sigmat}\frac{\de\sigmat}{\de\,\cos\theta^{*}_X} =
\frac{1}{2}(1+\pol\alpha_X\cos\theta^{*}_X) =
\left(\frac{1}{2}+A_X \cos\theta^{*}_X\right) ,
\label{eq:cosThetaDistr}
%\frac{d\Gamma}{d\,\cos{\theta_X}} = \frac{\Gamma}{2}(1+P_t\alpha_X\cos{\theta_X})\, ,
\end{equation}
where \pol is the single-top polarisation (due to the production vertex) along the top-quark spin direction, and $\alpha_X$ (known as ``spin-analysing power'') is a coefficient specific of the decay particle, very close to 100\% for the charged lepton~\cite{Jezabek:1994zv,Mahlon:1999gz}. Both \pol and $\alpha_X$ can be affected by top-quark anomalous couplings.

Single-top polarisation at the production vertex in the t-channel process is manifest in the slope of the \costheta\ distribution, where \thetaL\ is defined as the angle between the charged lepton and the light jet ($j^{\prime}$) in the reconstructed top quark rest frame. The rationale for this definition is that the (light) quark recoiling against the top quark tends to have a direction parallel to the spin direction of the top quark at the production vertex~\cite{Mahlon:1999gz}. The SM expectation $\pol\times\alpha_{\ell}\approx 1$ is part of the implicit assumptions of all the multivariate analyses at Tevatron and LHC and has even been built in the foundations of the earliest t-channel cross section measurements at the LHC~\cite{Chatrchyan:2011vp} which made use of the \costheta\ distribution to identify a SM-like signal~\footnote{This is named the ``2D analysis'' in reference~\cite{Chatrchyan:2011vp}, being based on a likelihood fit to the bi-dimensional distribution in the (\costheta,\etalj) plane.}. 
 A dedicated measurement of single-top polarisation has been performed by CMS with 8~TeV data~\cite{CMS-PAS-TOP-13-001}.
%, after a previous attempt at Tevatron, never published~\cite{cdf_polarization}.
 After a model-independent selection targeting t-channel production, the observed \costheta\ distribution is used to infer the differential cross section as a function of the parton-level \costheta. This is found to be compatible with the linear expectation of Eq.~\ref{eq:cosThetaDistr}, and a linear fit yields $\pol\times\alpha_{\ell} = 0.52 \pm 0.06~\rm{(stat)} \pm 0.20~\rm{(syst)}$, compatible with the SM expectation (0.88 according to different matrix-element generators interfaced to different parton-shower programs) within two Gaussian standard deviations.

%%% ATLAS Im(g_R)
%\somespace
 %The ATLAS experiment, using the 7~TeV data set, measures a forward-backward asymmetry with respect the normal to the plane defined by the W boson momentum and the top quark polarization~\cite{ATLAS-CONF-2013-032}.
% This measurement, together with a theoretical prediction (at NLO in QCD) of \pol, is used to constrain the $\cal{CP}$-violating imaginary component of the parameter $g_R$ in Eq.~\ref{eq:tbW}.

% Several angular observables can be useful, alone or combined, thanks to their sensitivities to the helicity states of the $W$ from top decay, to constrain the parameters of the Lagrangian of Eq.~\ref{eq:tbW}.

\somespace
The W-boson helicity fractions are sensitive to the anomalous couplings in Eq.~\ref{eq:tbW}. 
 The helicity angle \thetaW is defined as the angle between the W-boson momentum in the top-quark rest frame and the momentum of the down-type fermion from the W-boson decay, in the rest frame of the mother particle.
The functional form of the top quark partial decay width can be written as a function of the right-handed ($F_R$), left-handed ($F_L$) and longitudinal ($F_0$) helicity fractions of the W boson. 
 The CMS collaboration measured these decay widths in a phase-space region defined by a selection enriched in t-channel single top-quark events~\cite{Khachatryan:2014vma}, statistically independent from the measurements performed in \ttbar-optimised selections~\cite{CMS-PAS-TOP-12-025}.
 Similarly to the traditional \ttbar-based analyses, the measurement is based on the \costhetaW distribution. 
 Although the \ttbar and tW yields are reduced by the selection, that requires one charged lepton and exactly two jets (only one of them passing b-tagging selection), they are treated as a component of the signal as they carry information on the parameters of interest.
 Two out of three W-boson helicity fractions (as their sum is bound to unity) and the W+jets background contamination are treated as independent fit parameters.
 The results are found to be in good agreement with the \ttbar-optimised measurements, confirming the independence of the helicity fractions from the kinematic selection, and with NNLO standard model predictions.
 The helicity measurements are also interpreted to set limits on the real part of the anomalous couplings $g_L$ and $g_R$ of Eq.~\ref{eq:tbW}.

%\somespace
The ATLAS experiment, using the 7~TeV data set, performs an analysis of angular distributions of the decay products of single top quarks produced in the t channel~\cite{Aad:2015yem}. The doubly-differential normalised cross section as a function of \costhetaW and \phiW, where the latter is the complementary azimuthal angle to \thetaW defined as above, are used to extract the fraction $f_1$ of decays containing transversely polarised W bosons  and the phase $\delta_{-}$ between amplitudes for transversely and longitudinally polarised W bosons recoiling against
left-handed b quarks. The correlation in the measurement of these parameters is found to be 0.15. These values are translated into two-dimensional limits on the real and imaginary components of ratio of the coupling parameters $g_R$ and $f_L$ of Eq.~\ref{eq:tbW}.

\somespace
 Other angular observables can be defined in single top-quark events for the purpose of measuring top quark polarisation in a model-independent way (see, for example, ref.~\cite{AguilarSaavedra:2012xe}).

\section{Single top quark plus Higgs and the sign of \yt}
\label{sec:tHq}

While the most straightforward way to study the top quark Yukawa coupling (\yt) is through the measurement of top-anti-top quark production in association with a Higgs boson (\ttH), the interaction of the Higgs boson with the top quark can also be probed by studying the associated production of a single top quark and a Higgs boson~\cite{BORDES1993315}.
 What makes this signature special is its tree-level sensitivity to the relative phase between \yt and the coupling of the Higgs to the gauge bosons, as opposed to ${\rm H\to\gamma\gamma}$ and ${\rm gg\to HZ}$~\cite{Hespel:2015zea}, whose rates and distributions depend on the same phase through loop corrections and whose interpretation is therefore more model-dependent. The \ttH process is only sensitive to the modulus of \yt.

\begin{figure}[th!]
        \centering
        \includegraphics[width=0.3\textwidth]{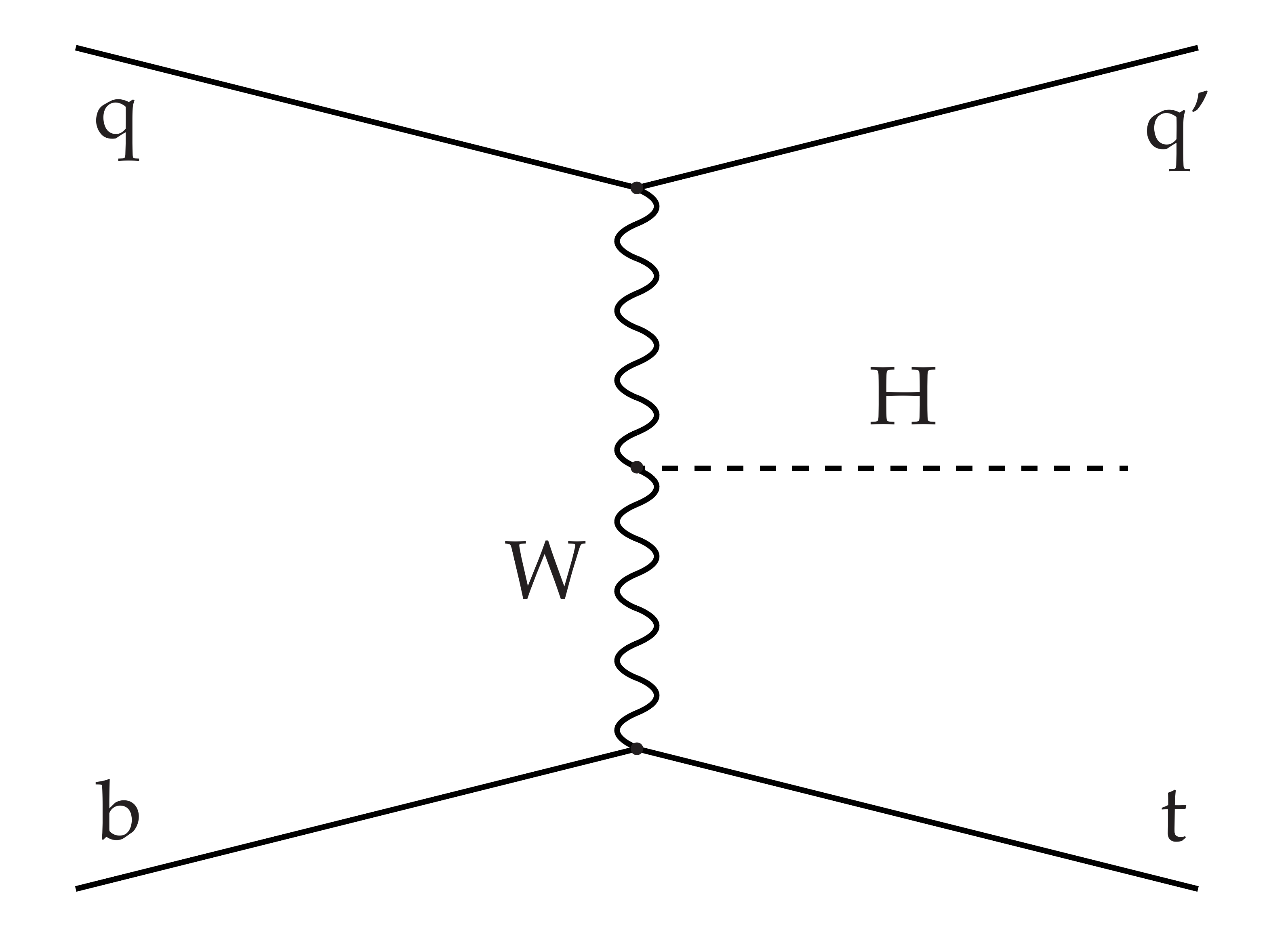}\hspace{1cm}
        \includegraphics[width=0.3\textwidth]{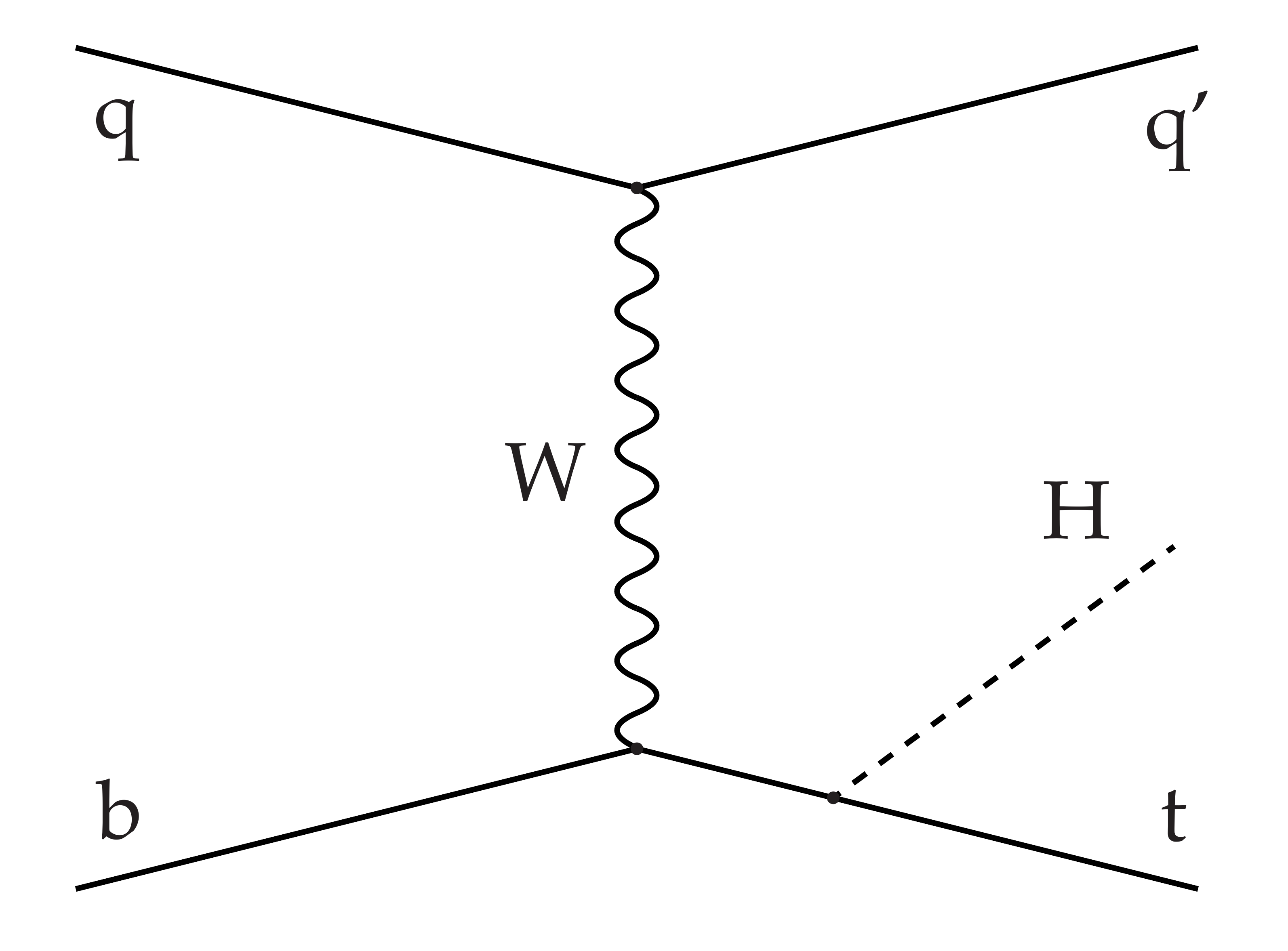}\vspace{.5cm}
        \caption{Dominant Feynman diagrams for the production of \tHq events.}
%: the Higgs boson is typically radiated
%        from the heavier legs of the diagram,  i.e. the W boson (left) or the top quark (right).}
\label{fig:thq_prod}
\end{figure}

 Single top quark plus Higgs boson production proceeds mainly through t-channel diagrams (\tHq), as in Fig.~\ref{fig:thq_prod}. 
The cross section of the \tHq process at 8~TeV is around 18~fb at NLO~\cite{Farina:2012xp}.
 As the couplings of the Higgs boson to the W boson and the top quark have opposite sign in the SM, these two diagrams interfere destructively, while any new physics inducing a phase between HWW and Htt couplings~\footnote{This phase can be an effective one, e.g. a two-Higgs doublets model, by adding a diagram with a charged Higgs boson exchanged in t channel, would mimic a complex phase in the coupling~\cite{Maltoni2001}.} would lead to an enhancement in this cross section, e.g. $\yt=-1$ would yield a cross section of 235~fb, significantly exceeding the \ttH cross section (130 fb)~\cite{Heinemeyer:2013tqa}. While the SM rate is arguably too low to be observed with available and future LHC data, the large enhancement for the flipped \yt sign potentially allows to observe or exclude this case, as has been suggested in a number of phenomenological papers (see, for example, \cite{Biswas:2012bd,Farina:2012xp,Biswas:2013xva,Chang:2014rfa}).

\begin{figure}[!htbp]
  {\centering
    \includegraphics[width=0.49\textwidth]{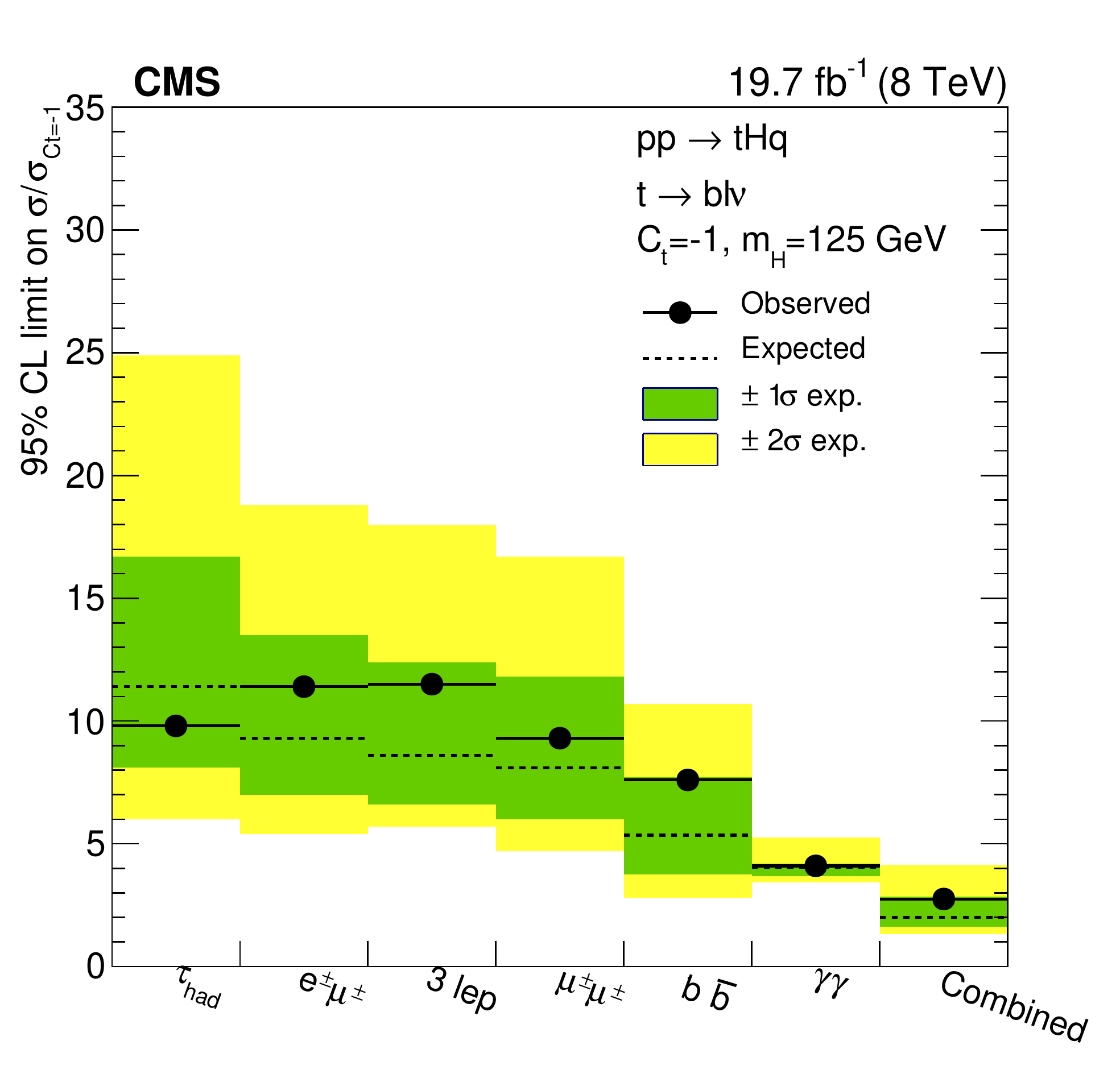}
    \includegraphics[width=0.49\textwidth]{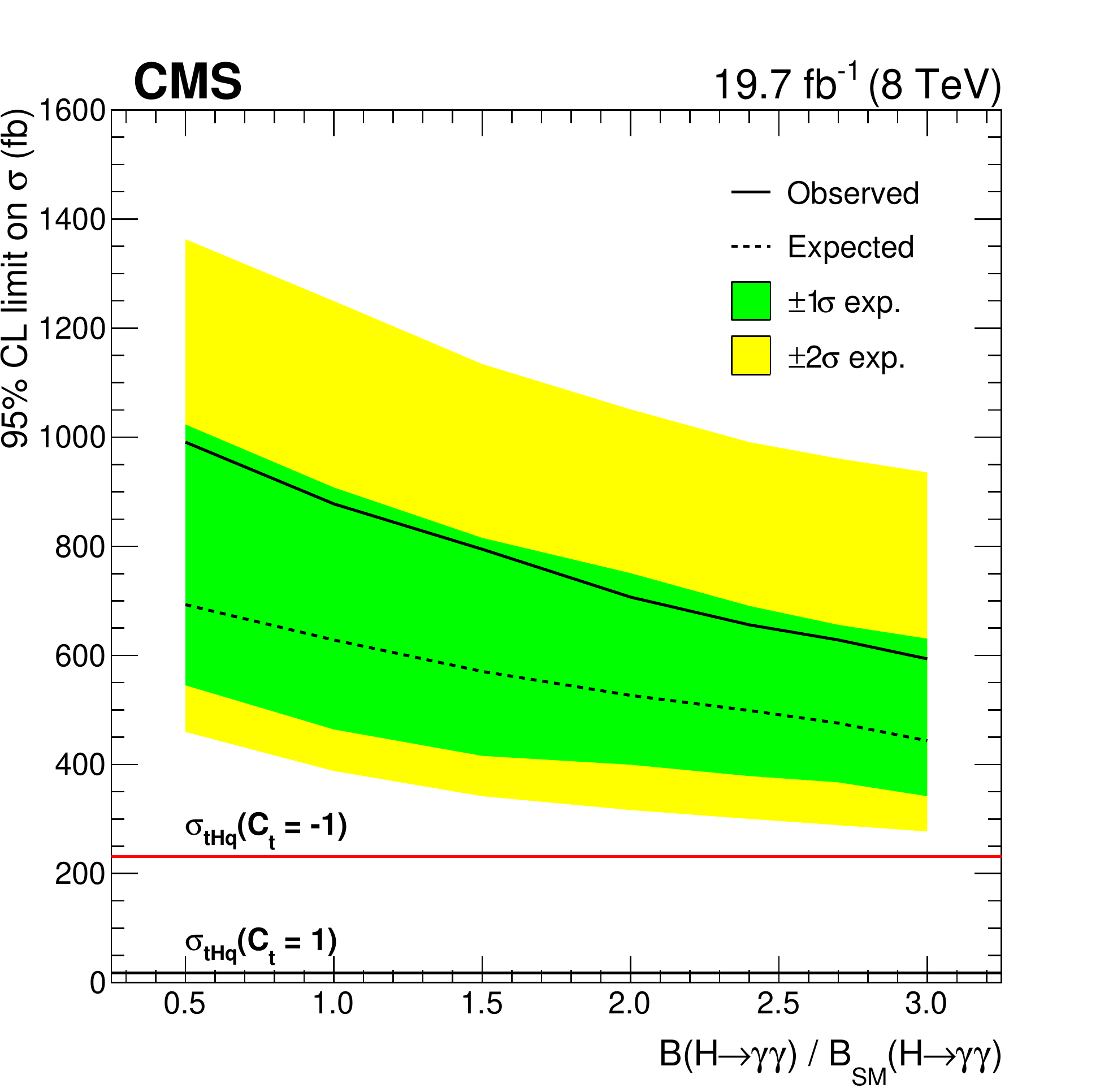}
  \caption{Left: 95\% CL upper limits on the excess event yields predicted by the enhanced \tHq and Higgs to di-photon branching fraction for $\yt=-1$ scenario, by decay channel and combined. Right: 95\% CL upper limits on the \tHq production cross section versus \BRHgg; the red horizontal line shows the predicted \tHq cross section for the SM Higgs boson with $m_{\rm H}$ = 125 GeV in the $\yt=-1$ scenario, while the black horizontal line shows the predicted \tHq cross section for the SM (i.e., $\yt=+1$) scenario. Figures from ref.~\cite{CMS-PAS-HIG-14-027}.
}
    \label{fig:thq_limits}}
\end{figure}

CMS performed dedicated searches for \tHq in a variety of signatures: $\gamma\gamma$, ${\rm b\bar b}$, same-sign leptons, three leptons, and electron or muon plus hadronically-decaying $\tau$~\cite{CMS-PAS-HIG-14-027}. In all Higgs decay channels, the top quark is assumed to decay semileptonically.
The data generally agree with the standard model expectations, and limits are set in the individual channels and combined
 with and without the assumption that the value of \yt affects \BRHgg and $\sigma_{\tHq}$ coherently. When this assumption is made, as in Fig.~\ref{fig:thq_limits} (left), the $\gamma\gamma$ channel is the most sensitive (as predicted in ref.~\cite{Biswas:2012bd}).
 The combined limit is also provided with \BRHgg treated as a free parameter, see Fig.~\ref{fig:thq_limits} (right).

The ATLAS collaboration followed a different approach~\cite{Aad:2014lma}. Instead of a direct search for this process, single top quark plus Higgs production is included in the signal model in a \ttH-optimised search in the ${\rm H\to\gamma\gamma}$ decay channel, which allows to set limits on negative values of \yt.

\section{Outlook: prospects at LHC in Run II}
\label{sec:outlook}

%The single-top final states are an important place to look for indirect evidences of new heavy quarks, and the measurements of the properties of single top quarks are crucial to gain sensitivity to non-SM interactions and to get some insight into the kind of new physics at hand in case of deviations from the SM hypothesis.

%\somespace
The first inclusive cross-section measurements for t-channel production are expected very soon with the early 13~TeV data collected by ATLAS and CMS during 2015.
 With an expected jump by a factor of 2.6~\cite{Kant:2014oha} in rate with respect to 8~TeV collisions, this production mode will soon allow very extended differential studies as a function of many observables. The tW process, discovered at the end of Run I, will enter the realm of ``precision physics'' in Run II thanks to an increase by a factor 3.4 in rate~\cite{Kidonakis:2013zqa}.

% A possibility not yet exploited by the LHC analyses is to take advantage of the different dependencies of these processes on energy, e.g., by simultaneous fits to 7 and 8~TeV data with $\sigma_{bkg}({\rm 8~TeV})/\sigma_{bkg}({\rm 7~TeV})$ ratios constrained to SM expectations.

The price to pay for increased precision is that old approximations become more and more inadequate.
 In the single-top sector, the definition of the signal itself is challenged, as the old rigid separation of the three canonical SM processes among them and with their backgrounds, inspired by their Born-level diagrams, does not hold water when higher-order effects are considered.
 An egregious example is the tW production mode, whose interference at NLO with its main background, \ttbar, generated a long literature. 
 As a consequence, increasing attention is being paid in this community to the definition of proper fiducial regions of the phase space, such that unambiguous interpretations can be extracted from the results~\footnote{ATLAS reported a fiducial cross section measurement in the ${\rm tW+}$\ttbar signature in ref.~\cite{Aad:2015eto}. ATLAS and CMS also reported fiducial cross section measurements in the t-channel production mode, but only preliminary results are reported at the time of writing.}.

 With more statistics becoming available in Run II, the author modestly suggests to specifically address phase space regions where modelling is most problematic, as those can give the most valuable insight. Monte Carlo models of the t-channel mode have always struggled to provide a satisfactory description of the kinematic properties of the associated b quark, and different simulations schemes (e.g., ``4-flavour'' versus ``5-flavour'' scheme~\cite{fourfiveflavorschemesmalt,fourfiveflavorschemescamp}, a distinction based on which quark flavours are treated as sea-quark components of the protons) are known to yield different predictions, that can be challenged by precise cross section measurements as a function of $p_T^{\rm b}$, $y^{\rm b}$.
 Similarly, it would be very interesting to address the issue of tW/\ttbar interference by measuring the properties of top quarks in portions of the phase space with enhanced sensitivity to this effect.
 Until this kind of studies are performed we will continue to need large ``theory systematics'' assessed by comparing different simulations approaches, none of them fully theoretically sound.

More data are not guaranteed to make the study of the s-channel process easier, as the current analyses are limited by systematic uncertainties.
 The signal cross section at 13~TeV is only twice than at 8~TeV~\cite{Kant:2014oha} while the dominant background, \ttbar, is 3.3 times larger~\cite{Czakon:2011xx}. %As the current analyses are already limited by systematic uncertainties, measuring s-channel single top at LHC would require a significant progress on the theory side such as to reduce the signal and background modelling uncertainties, and new ideas for experimental break-through.

The next single-top process to be observed may be Z-associated production, despite the small cross section in the standard model (roughly half picobarn at 13~TeV~\cite{Campbell:2013yla}, to be compared for example with 10~pb for the s-channel mode~\cite{Kant:2014oha}), thanks to the very clean signature of three leptons, two of them forming an invariant mass around $m_{\rm Z}$, and one b-tagged jet.
 The same signature is a powerful probe of FCNC interactions, complementary to searches for ${\rm t\to Zq}$ decay in \ttbar events~\cite{Abazov:2011qf,Aaltonen:2008ac,Aad:2012ij,Chatrchyan:2013nwa}.

The author expects the search for H-associated production to attract more attention in Run II.
 Its cross section is expected to increase by a factor 4~\cite{Demartin:2015uha}, independently from the value of \yt.
 A naive extrapolation can be made of the current upper limit in the ${\rm H\to\gamma\gamma}$ channel~\cite{CMS-PAS-HIG-14-027}, as this channel is by far statistics-limited and practically background free. Assuming the selection efficiency to be similar to the 8~TeV analysis, observing 0 event counts with the first 20~\fb would rule out the $\yt=-1$ value with this channel alone~\footnote{Assuming negligible systematics and negligible backgrounds, the Poissonian probability to observe no events is $\exp(-\epsilon\sigma\cal{L})$, giving an upper limit at 95\% confidence level of $\sigma_{95\%~limit}-\log(0.05)/\epsilon\cal{L}$. The $y_t=-1$ case would be excluded if $\sigma_{95\%~limit}/\sigma_{y_t=-1}<1$. Assuming the selection efficiency $\epsilon$ to be the same, the predicted increase of a factor 4 in $\sigma_{y_t=-1}$ implies that 1/4 of the integrated luminosity $\cal{L}$ is needed to set the same upper limit, and the same luminosity to exclude $y_t=-1$.}.

\somespace
Concluding, the LHC Run I brought single-top studies into the domain of precision physics; Run II may bring it into the domain of beyond-SM discovery.

\section*{Acknowledgements}

The author acknowledges Nikolaos Kidonakis for his kindness in providing the theory curves in Figure~\ref{fig:sqrts} and Alberto Orso Maria Iorio, Luca Lista, Rebeca Gonzalez Suarez for having contributed to earlier versions of that figure.
Figure~\ref{fig:vtb} was created in the context of the LHC Top Working Group~\cite{toplhcwg}, with direct or indirect contributions by Abideh (Nadjieh) Jafari, Alberto Orso Maria Iorio, Julien Donini, Luca Lista, Matthias Komm, Oliver Maria Kind, and Reinhard Schwienhorst.
 The author is also indebted to Andrey Popov, Fabio Maltoni, Andrea Thamm and Emidio Gabrielli for several discussions about \tHq.
 %The author wishes to thank Andreas Meyer, FIXME and FIXME for their comments to the preliminary drafts of this paper.

%\somespace
%{\small
%{\bf Temporary note:}
 %As a general rule here I only cite papers, not "CMS PAS" or "ATLAS CONF-NOTE", as these are "preliminary results" expected to be superseded eventually by a refereed publication.
% I am trying to stick to this rule, and for example this is why I am not mentioning several nice ATLAS and CMS measurements.
%measurements (e.g. top mass in single top topologies, fiducial cross section in t channel, tW at 8~TeV, $Im(g_R)$ extraction from an angular observable, s-channel limit at 7~TeV), or the t-channel differential cross section by CMS.
% Exceptions: combinations by the Top LHC Working Group~\footnote{By policy, these are never submitted to a journal (or to arXiv), and are only superseded by newer combinations}
%and CMS PAS that have a fair chance to be on arXiv by the end of the review for this paper, e.g.~\cite{CMS-PAS-TOP-13-009,CMS-PAS-TOP-14-003,CMS-PAS-TOP-13-001}.
%\item CMS PAS that are not going to ever become a publication as their main authors left, e.g.~\cite{CMS-PAS-TOP-14-004} (by being one of the conveners of the relevant physics group, I have insider information about that);
%\item ATLAS CONF-NOTE that are old enough that I am assuming that they are not going to become a publication, e.g.~\cite{ATLAS-CONF-2013-032} (same applies to Tevatron, e.g.~\cite{cdf_polarization}); please let me know if you prefer them to be removed.
%\end{itemize}
%}

\small
\bibliographystyle{cms-note}
\bibliography{biblio}

\providecommand{\href}[2]{#2}\begingroup\raggedright\begin{thebibliography}{100}%
\makeatletter
\providecommand{\hrefCMSnoop }[0]{\@secondoftwo}%
\makeatother
\providecommand{\doi}{\texttt{doi:}\begingroup \urlstyle{tt}\Url}

\bibitem{Abe:1995hr}
\hrefCMSnoop {} {{ CDF} Collaboration, ``{Observation of top quark production
  in $\rm \bar{p}p$ collisions}'',} \textit{ Phys. Rev. Lett.} \textbf{ 74}
  (1995) 2626--2631,
  \href{http://dx.doi.org/10.1103/PhysRevLett.74.2626}{\doi{10.1103/PhysRevLett.74.2626}},
\href{http://www.arXiv.org/abs/hep-ex/9503002}{\texttt{ arXiv:hep-ex/9503002}}.
%%CITATION = HEP-EX/9503002;%%.

\bibitem{Abachi:1995iq}
\hrefCMSnoop {} {{ D0} Collaboration, ``{Observation of the top quark}'',}
  \textit{ Phys. Rev. Lett.} \textbf{ 74} (1995) 2632--2637,
  \href{http://dx.doi.org/10.1103/PhysRevLett.74.2632}{\doi{10.1103/PhysRevLett.74.2632}},
\href{http://www.arXiv.org/abs/hep-ex/9503003}{\texttt{ arXiv:hep-ex/9503003}}.
%%CITATION = HEP-EX/9503003;%%.

\bibitem{Abazov:2009ii}
\hrefCMSnoop {} {{ D0} Collaboration, ``{Observation of Single Top-Quark
  Production}'',} \textit{ Phys. Rev. Lett.} \textbf{ 103} (2009) 092001,
  \href{http://dx.doi.org/10.1103/PhysRevLett.103.092001}{\doi{10.1103/PhysRevLett.103.092001}},
\href{http://www.arXiv.org/abs/0903.0850}{\texttt{ arXiv:0903.0850}}.
%%CITATION = 0903.0850;%%.

\bibitem{Aaltonen:2009jj}
\hrefCMSnoop {} {{ CDF} Collaboration, ``{First Observation of Electroweak
  Single Top Quark Production}'',} \textit{ Phys. Rev. Lett.} \textbf{ 103}
  (2009) 092002,
  \href{http://dx.doi.org/10.1103/PhysRevLett.103.092002}{\doi{10.1103/PhysRevLett.103.092002}},
\href{http://www.arXiv.org/abs/0903.0885}{\texttt{ arXiv:0903.0885}}.
%%CITATION = 0903.0885;%%.

\bibitem{Aad:2008zzm}
\hrefCMSnoop {} {{ ATLAS} Collaboration, ``{The ATLAS Experiment at the CERN
  Large Hadron Collider}'',} \textit{ JINST} \textbf{ 3} (2008) S08003,
\href{http://dx.doi.org/10.1088/1748-0221/3/08/S08003}{\doi{10.1088/1748-0221/3/08/S08003}}.
%%CITATION = JINST,3,S08003;%%.

\bibitem{Chatrchyan:2008aa}
\hrefCMSnoop {} {{ CMS} Collaboration, ``{The CMS experiment at the CERN
  LHC}'',} \textit{ JINST} \textbf{ 3} (2008) S08004,
\href{http://dx.doi.org/10.1088/1748-0221/3/08/S08004}{\doi{10.1088/1748-0221/3/08/S08004}}.
%%CITATION = JINST,3,S08004;%%.

\bibitem{Chatrchyan:2011vp}
\hrefCMSnoop {} {{ CMS} Collaboration, ``{Measurement of the t-channel single
  top quark production cross section in pp collisions at $\sqrt{s} =
  7$~TeV}'',} \textit{ Phys.Rev.Lett.} \textbf{ 107} (2011) 091802,
  \href{http://dx.doi.org/10.1103/PhysRevLett.107.091802}{\doi{10.1103/PhysRevLett.107.091802}},
\href{http://www.arXiv.org/abs/1106.3052}{\texttt{ arXiv:1106.3052}}.
%%CITATION = ARXIV:1106.3052;%%.

\bibitem{Aad:2012ux}
\hrefCMSnoop {} {{ ATLAS} Collaboration, ``{Measurement of the t-channel single
  top-quark production cross section in pp collisions at $\sqrt{s}=7$ TeV with
  the ATLAS detector}'',} \textit{ Physics Letters B} \textbf{ 717} (2012)
  330--350,
\href{http://www.arXiv.org/abs/1205.3130}{\texttt{ arXiv:1205.3130}}.
%%CITATION = ARXIV:1205.3130;%%.

\bibitem{Tait:2000sh}
\hrefCMSnoop {} {T.~M.~P. Tait and C.~P. Yuan, ``{Single top quark production
  as a window to physics beyond the standard model}'',} \textit{ Phys. Rev.}
  \textbf{ D63} (2000) 014018,
  \href{http://dx.doi.org/10.1103/PhysRevD.63.014018}{\doi{10.1103/PhysRevD.63.014018}},
\href{http://www.arXiv.org/abs/hep-ph/0007298}{\texttt{ arXiv:hep-ph/0007298}}.
%%CITATION = HEP-PH/0007298;%%.

\bibitem{Aad:2012ej}
\hrefCMSnoop {} {{ ATLAS} Collaboration, ``{Search for tb resonances in
  proton-proton collisions at $\sqrt{s}=7$ TeV with the ATLAS detector}'',}
  \textit{ Phys. Rev. Lett.} \textbf{ 109} (2012) 081801,
  \href{http://dx.doi.org/10.1103/PhysRevLett.109.081801}{\doi{10.1103/PhysRevLett.109.081801}},
\href{http://www.arXiv.org/abs/1205.1016}{\texttt{ arXiv:1205.1016}}.
%%CITATION = ARXIV:1205.1016;%%.

\bibitem{Aad:2014xra}
\hrefCMSnoop {} {{ ATLAS} Collaboration, ``{Search for $\rm W' \rightarrow tb
  \rightarrow qqbb$ decays in pp collisions at $\sqrt{s}$ = 8 TeV with the
  ATLAS detector}'',} \textit{ Eur. Phys. J.} \textbf{ C75} (2015), no.~4, 165,
  \href{http://dx.doi.org/10.1140/epjc/s10052-015-3372-2}{\doi{10.1140/epjc/s10052-015-3372-2}},
\href{http://www.arXiv.org/abs/1408.0886}{\texttt{ arXiv:1408.0886}}.
%%CITATION = ARXIV:1408.0886;%%.

\bibitem{Aad:2014xea}
\hrefCMSnoop {} {{ ATLAS} Collaboration, ``{Search for $\rm W' \to t\bar{b}$ in
  the lepton plus jets final state in proton-proton collisions at a
  centre-of-mass energy of $\sqrt{s}$ = 8 TeV with the ATLAS detector}'',}
  \textit{ Phys. Lett.} \textbf{ B743} (2015) 235--255,
  \href{http://dx.doi.org/10.1016/j.physletb.2015.02.051}{\doi{10.1016/j.physletb.2015.02.051}},
\href{http://www.arXiv.org/abs/1410.4103}{\texttt{ arXiv:1410.4103}}.
%%CITATION = ARXIV:1410.4103;%%.

\bibitem{Chatrchyan:2014koa}
\hrefCMSnoop {} {{ CMS} Collaboration, ``{Search for W' $\to $ tb decays in the
  lepton + jets final state in pp collisions at $\sqrt{s}$ = 8 TeV}'',}
  \textit{ JHEP} \textbf{ 05} (2014) 108,
  \href{http://dx.doi.org/10.1007/JHEP05(2014)108}{\doi{10.1007/JHEP05(2014)108}},
\href{http://www.arXiv.org/abs/1402.2176}{\texttt{ arXiv:1402.2176}}.
%%CITATION = ARXIV:1402.2176;%%.

\bibitem{Khachatryan:2015edz}
\hrefCMSnoop {} {{ CMS} Collaboration, ``{Search for W' $\to$ tb in
  proton-proton collisions at $\sqrt{s} = $ 8 TeV}'',}
\href{http://www.arXiv.org/abs/1509.06051}{\texttt{ arXiv:1509.06051}}.
%%CITATION = ARXIV:1509.06051;%%.

\bibitem{Aad:2013rna}
\hrefCMSnoop {} {{ ATLAS} Collaboration, ``{Search for single $\rm b^*$-quark
  production with the ATLAS detector at $\sqrt{s}=7$ TeV}'',} \textit{ Phys.
  Lett.} \textbf{ B721} (2013) 171--189,
  \href{http://dx.doi.org/10.1016/j.physletb.2013.03.016}{\doi{10.1016/j.physletb.2013.03.016}},
\href{http://www.arXiv.org/abs/1301.1583}{\texttt{ arXiv:1301.1583}}.
%%CITATION = ARXIV:1301.1583;%%.

\bibitem{Aad:2014efa}
\hrefCMSnoop {} {{ ATLAS} Collaboration, ``{Search for pair and single
  production of new heavy quarks that decay to a Z boson and a third-generation
  quark in pp collisions at $\sqrt{s}=8$ TeV with the ATLAS detector}'',}
  \textit{ JHEP} \textbf{ 11} (2014) 104,
  \href{http://dx.doi.org/10.1007/JHEP11(2014)104}{\doi{10.1007/JHEP11(2014)104}},
\href{http://www.arXiv.org/abs/1409.5500}{\texttt{ arXiv:1409.5500}}.
%%CITATION = ARXIV:1409.5500;%%.

\bibitem{Aad:2015voa}
\hrefCMSnoop {} {{ ATLAS} Collaboration, ``{Search for the production of single
  vector-like and excited quarks in the Wt final state in pp collisions at
  $\sqrt{s}$ = 8 TeV with the ATLAS detector}'',}
\href{http://www.arXiv.org/abs/1510.02664}{\texttt{ arXiv:1510.02664}}.
%%CITATION = ARXIV:1510.02664;%%.

\bibitem{Khachatryan:2015mta}
\hrefCMSnoop {} {{ CMS} Collaboration, ``{Search for the Production of an
  Excited Bottom Quark Decaying to tW in Proton-Proton Collisions at $\sqrt{s}$
  = 8 TeV}'',}
\href{http://www.arXiv.org/abs/1509.08141}{\texttt{ arXiv:1509.08141}}.
%%CITATION = ARXIV:1509.08141;%%.

\bibitem{Aad:2014wza}
\hrefCMSnoop {} {{ ATLAS} Collaboration, ``{Search for invisible particles
  produced in association with single-top-quarks in proton-proton collisions at
  $\sqrt{s}$ = 8 TeV with the ATLAS detector}'',} \textit{ Eur. Phys. J.}
  \textbf{ C75} (2015), no.~2, 79,
  \href{http://dx.doi.org/10.1140/epjc/s10052-014-3233-4}{\doi{10.1140/epjc/s10052-014-3233-4}},
\href{http://www.arXiv.org/abs/1410.5404}{\texttt{ arXiv:1410.5404}}.
%%CITATION = ARXIV:1410.5404;%%.

\bibitem{Khachatryan:2014uma}
\hrefCMSnoop {} {{ CMS} Collaboration, ``{Search for Monotop Signatures in
  Proton-Proton Collisions at $\sqrt s =$ 8 TeV}'',} \textit{ Phys. Rev. Lett.}
  \textbf{ 114} (2015), no.~10, 101801,
  \href{http://dx.doi.org/10.1103/PhysRevLett.114.101801}{\doi{10.1103/PhysRevLett.114.101801}},
\href{http://www.arXiv.org/abs/1410.1149}{\texttt{ arXiv:1410.1149}}.
%%CITATION = ARXIV:1410.1149;%%.

\bibitem{Kidonakis:2011wy}
\hrefCMSnoop {} {N.~Kidonakis, ``{Next-to-next-to-leading-order collinear and
  soft gluon corrections for t-channel single top quark production}'',}
  \textit{ Phys. Rev.} \textbf{ D83} (2011) 091503,
  \href{http://dx.doi.org/10.1103/PhysRevD.83.091503}{\doi{10.1103/PhysRevD.83.091503}},
\href{http://www.arXiv.org/abs/1103.2792}{\texttt{ arXiv:1103.2792}}.
%%CITATION = ARXIV:1103.2792;%%.

\bibitem{Kidonakis:2010ux}
\hrefCMSnoop {} {N.~Kidonakis, ``{Two-loop soft anomalous dimensions for single
  top quark associated production with a $\rm W^-$ or $\rm H^-$}'',} \textit{
  Phys. Rev.} \textbf{ D82} (2010) 054018,
  \href{http://dx.doi.org/10.1103/PhysRevD.82.054018}{\doi{10.1103/PhysRevD.82.054018}},
\href{http://www.arXiv.org/abs/1005.4451}{\texttt{ arXiv:1005.4451}}.
%%CITATION = ARXIV:1005.4451;%%.

\bibitem{Kidonakis:2010tc}
\hrefCMSnoop {} {N.~Kidonakis, ``{NNLL resummation for s-channel single top
  quark production}'',} \textit{ Phys. Rev.} \textbf{ D81} (2010) 054028,
  \href{http://dx.doi.org/10.1103/PhysRevD.81.054028}{\doi{10.1103/PhysRevD.81.054028}},
\href{http://www.arXiv.org/abs/1001.5034}{\texttt{ arXiv:1001.5034}}.
%%CITATION = ARXIV:1001.5034;%%.

\bibitem{Brucherseifer:2014ama}
\hrefCMSnoop {} {M.~Brucherseifer, F.~Caola, and K.~Melnikov, ``{On the NNLO
  QCD corrections to single-top production at the LHC}'',} \textit{ Phys.
  Lett.} \textbf{ B736} (2014) 58--63,
  \href{http://dx.doi.org/10.1016/j.physletb.2014.06.075}{\doi{10.1016/j.physletb.2014.06.075}},
\href{http://www.arXiv.org/abs/1404.7116}{\texttt{ arXiv:1404.7116}}.
%%CITATION = ARXIV:1404.7116;%%.

\bibitem{Aliev:2010zk}
M.~Aliev\hrefCMSnoop {} { {et~al.}, ``{HATHOR: HAdronic Top and Heavy quarks
  crOss section calculatoR}'',} \textit{ Comput. Phys. Commun.} \textbf{ 182}
  (2011) 1034--1046,
  \href{http://dx.doi.org/10.1016/j.cpc.2010.12.040}{\doi{10.1016/j.cpc.2010.12.040}},
\href{http://www.arXiv.org/abs/1007.1327}{\texttt{ arXiv:1007.1327}}.
%%CITATION = ARXIV:1007.1327;%%.

\bibitem{Kant:2014oha}
P.~Kant\hrefCMSnoop {} { {et~al.}, ``{HatHor for single top-quark production:
  Updated predictions and uncertainty estimates for single top-quark production
  in hadronic collisions}'',} \textit{ Comput. Phys. Commun.} \textbf{ 191}
  (2015) 74--89,
  \href{http://dx.doi.org/10.1016/j.cpc.2015.02.001}{\doi{10.1016/j.cpc.2015.02.001}},
\href{http://www.arXiv.org/abs/1406.4403}{\texttt{ arXiv:1406.4403}}.
%%CITATION = ARXIV:1406.4403;%%.

\bibitem{Aaltonen:2015cra}
\hrefCMSnoop {} {{ CDF, D0} Collaboration, ``{Tevatron combination of
  single-top-quark cross sections and determination of the magnitude of the
  Cabibbo-Kobayashi-Maskawa matrix element $V_{\rm tb}$}'',}
\href{http://www.arXiv.org/abs/1503.05027}{\texttt{ arXiv:1503.05027}}.
%%CITATION = ARXIV:1503.05027;%%.

\bibitem{Aad:2014fwa}
\hrefCMSnoop {} {{ ATLAS} Collaboration, ``{Comprehensive measurements of
  t-channel single top-quark production cross sections at $\sqrt{s} = 7$ TeV
  with the ATLAS detector}'',} \textit{ Phys. Rev.} \textbf{ D90} (2014),
  no.~11, 112006,
  \href{http://dx.doi.org/10.1103/PhysRevD.90.112006}{\doi{10.1103/PhysRevD.90.112006}},
\href{http://www.arXiv.org/abs/1406.7844}{\texttt{ arXiv:1406.7844}}.
%%CITATION = ARXIV:1406.7844;%%.

\bibitem{Chatrchyan:2012ep}
\hrefCMSnoop {} {{ CMS} Collaboration, ``{Measurement of the single-top-quark
  t-channel cross section in pp collisions at $\sqrt{s}=7$ TeV}'',} \textit{
  JHEP} \textbf{ 1212} (2012) 035,
  \href{http://dx.doi.org/10.1007/JHEP12(2012)035}{\doi{10.1007/JHEP12(2012)035}},
\href{http://www.arXiv.org/abs/1209.4533}{\texttt{ arXiv:1209.4533}}.
%%CITATION = ARXIV:1209.4533;%%.

\bibitem{Khachatryan:2014iya}
\hrefCMSnoop {} {{ CMS} Collaboration, ``{Measurement of the t-channel
  single-top-quark production cross section and of the $V_{tb}$ CKM matrix
  element in pp collisions at $\sqrt{s}$= 8 TeV}'',} \textit{ JHEP} \textbf{
  06} (2014) 090,
  \href{http://dx.doi.org/10.1007/JHEP06(2014)090}{\doi{10.1007/JHEP06(2014)090}},
\href{http://www.arXiv.org/abs/1403.7366}{\texttt{ arXiv:1403.7366}}.
%%CITATION = ARXIV:1403.7366;%%.

\bibitem{ABM11}
\hrefCMSnoop {} {S.~Alekhin, J.~Bl{\"u}mlein, and S.~Moch, ``Parton
  distribution functions and benchmark cross sections at {NNLO}'',} \textit{
  Phys. Rev. D} \textbf{ 86} (2012) 0054009,
  \href{http://dx.doi.org/10.1103/PhysRevD.86.054009}{\doi{10.1103/PhysRevD.86.054009}},
  \href{http://www.arXiv.org/abs/1202.2281}{\texttt{ arXiv:1202.2281}}.

\bibitem{CT10}
H.-L. Lai\hrefCMSnoop {} { {et~al.}, ``{New parton distributions for collider
  physics}'',} \textit{ Phys. Rev. D} \textbf{ 82} (2010) 074024,
  \href{http://dx.doi.org/10.1103/PhysRevD.82.074024}{\doi{10.1103/PhysRevD.82.074024}},
  \href{http://www.arXiv.org/abs/1007.2241}{\texttt{ arXiv:1007.2241}}.

\bibitem{GJR08}
\hrefCMSnoop {} {M.~Gluck, P.~Jimenez-Delgado, E.~Reya, and C.~Schuck, ``{On
  the role of heavy flavor parton distributions at high energy colliders}'',}
  \textit{ Phys. Lett.} \textbf{ B664} (2008) 133--138,
  \href{http://dx.doi.org/10.1016/j.physletb.2008.04.063}{\doi{10.1016/j.physletb.2008.04.063}},
\href{http://www.arXiv.org/abs/0801.3618}{\texttt{ arXiv:0801.3618}}.
%%CITATION = ARXIV:0801.3618;%%.

\bibitem{HERAPDF}
\hrefCMSnoop {} {{H1 and ZEUS Collaborations}, ``{Combined measurement and QCD
  analysis of the inclusive ${e}^{\pm}{p}$ scattering cross sections at
  HERA}'',} \textit{ JHEP} \textbf{ 01} (2010) 109,
  \href{http://dx.doi.org/10.1007/JHEP01(2010)109}{\doi{10.1007/JHEP01(2010)109}},
  \href{http://www.arXiv.org/abs/0911.0884}{\texttt{ arXiv:0911.0884}}.

\bibitem{MSTW2008NLO}
\hrefCMSnoop {} {W.~J. Martin, A. D~Stirling and G.~Watt, ``{Parton
  distributions for the LHC}'',} \textit{ Eur. Phys. J. C} \textbf{ 63} (2009)
  189,
  \href{http://dx.doi.org/10.1140/epjc/s10052-009-1072-5}{\doi{10.1140/epjc/s10052-009-1072-5}}.

\bibitem{NNPDF}
R.~D. Ball\hrefCMSnoop {} { {et~al.}, ``{Parton distributions with LHC
  data}'',} \textit{ Nucl. Phys. B} \textbf{ 867} (2013) 244,
  \href{http://dx.doi.org/10.1016/j.nuclphysb.2012.10.003}{\doi{10.1016/j.nuclphysb.2012.10.003}},
  \href{http://www.arXiv.org/abs/1207.1303}{\texttt{ arXiv:1207.1303}}.

\bibitem{Aad:2012xca}
\hrefCMSnoop {} {{ ATLAS} Collaboration, ``{Evidence for the associated
  production of a W boson and a top quark in ATLAS at $\sqrt{s}=7$ TeV}'',}
  \textit{ Phys. Lett.} \textbf{ B716} (2012) 142--159,
  \href{http://dx.doi.org/10.1016/j.physletb.2012.08.011}{\doi{10.1016/j.physletb.2012.08.011}},
\href{http://www.arXiv.org/abs/1205.5764}{\texttt{ arXiv:1205.5764}}.
%%CITATION = ARXIV:1205.5764;%%.

\bibitem{Chatrchyan:2012zca}
\hrefCMSnoop {} {{ CMS} Collaboration, ``{Evidence for associated production of
  a single top quark and W boson in pp collisions at $\sqrt{s}$ = 7 TeV}'',}
  \textit{ Phys. Rev. Lett.} \textbf{ 110} (2013) 022003,
  \href{http://dx.doi.org/10.1103/PhysRevLett.110.022003}{\doi{10.1103/PhysRevLett.110.022003}},
\href{http://www.arXiv.org/abs/1209.3489}{\texttt{ arXiv:1209.3489}}.
%%CITATION = ARXIV:1209.3489;%%.

\bibitem{Chatrchyan:2014tua}
\hrefCMSnoop {} {{ CMS} Collaboration, ``{Observation of the associated
  production of a single top quark and a W boson in pp collisions at $\sqrt{s}$
  = 8 TeV}'',} \textit{ Phys.Rev.Lett.} \textbf{ 112} (2014) 231802,
  \href{http://dx.doi.org/10.1103/PhysRevLett.112.231802}{\doi{10.1103/PhysRevLett.112.231802}},
\href{http://www.arXiv.org/abs/1401.2942}{\texttt{ arXiv:1401.2942}}.
%%CITATION = ARXIV:1401.2942;%%.

\bibitem{Aad:2015eto}
\hrefCMSnoop {} {{ ATLAS} Collaboration, ``{Measurement of the production
  cross-section of a single top quark in association with a W boson at 8 TeV
  with the ATLAS experiment}'',}
\href{http://www.arXiv.org/abs/1510.03752}{\texttt{ arXiv:1510.03752}}.
%%CITATION = ARXIV:1510.03752;%%.

\bibitem{Belyaev:2000me}
\hrefCMSnoop {} {A.~Belyaev and E.~Boos, ``{Single top quark tW + X production
  at the CERN LHC: A Closer look}'',} \textit{ Phys. Rev.} \textbf{ D63} (2001)
  034012,
  \href{http://dx.doi.org/10.1103/PhysRevD.63.034012}{\doi{10.1103/PhysRevD.63.034012}},
\href{http://www.arXiv.org/abs/hep-ph/0003260}{\texttt{ arXiv:hep-ph/0003260}}.
%%CITATION = HEP-PH/0003260;%%.

\bibitem{Campbell:2005bb}
\hrefCMSnoop {} {J.~M. Campbell and F.~Tramontano, ``{Next-to-leading order
  corrections to Wt production and decay}'',} \textit{ Nucl. Phys.} \textbf{
  B726} (2005) 109--130,
  \href{http://dx.doi.org/10.1016/j.nuclphysb.2005.08.015}{\doi{10.1016/j.nuclphysb.2005.08.015}},
\href{http://www.arXiv.org/abs/hep-ph/0506289}{\texttt{ arXiv:hep-ph/0506289}}.
%%CITATION = HEP-PH/0506289;%%.

\bibitem{Frixione:2008yi}
S.~Frixione\hrefCMSnoop {} { {et~al.}, ``{Single-top hadroproduction in
  association with a W boson}'',} \textit{ JHEP} \textbf{ 07} (2008) 029,
  \href{http://dx.doi.org/10.1088/1126-6708/2008/07/029}{\doi{10.1088/1126-6708/2008/07/029}},
\href{http://www.arXiv.org/abs/0805.3067}{\texttt{ arXiv:0805.3067}}.
%%CITATION = ARXIV:0805.3067;%%.

\bibitem{Frixione:2002ik}
\hrefCMSnoop {} {S.~Frixione and B.~R. Webber, ``{Matching NLO QCD computations
  and parton shower simulations}'',} \textit{ JHEP} \textbf{ 06} (2002) 029,
  \href{http://dx.doi.org/10.1088/1126-6708/2002/06/029}{\doi{10.1088/1126-6708/2002/06/029}},
\href{http://www.arXiv.org/abs/hep-ph/0204244}{\texttt{ arXiv:hep-ph/0204244}}.
%%CITATION = HEP-PH/0204244;%%.

\bibitem{Frixione:2007vw}
\hrefCMSnoop {} {S.~Frixione, P.~Nason, and C.~Oleari, ``{Matching NLO QCD
  computations with Parton Shower simulations: the POWHEG method}'',} \textit{
  JHEP} \textbf{ 11} (2007) 070,
  \href{http://dx.doi.org/10.1088/1126-6708/2007/11/070}{\doi{10.1088/1126-6708/2007/11/070}},
\href{http://www.arXiv.org/abs/0709.2092}{\texttt{ arXiv:0709.2092}}.
%%CITATION = ARXIV:0709.2092;%%.

\bibitem{White:2009yt}
\hrefCMSnoop {} {C.~D. White, S.~Frixione, E.~Laenen, and F.~Maltoni,
  ``{Isolating Wt production at the LHC}'',} \textit{ JHEP} \textbf{ 11} (2009)
  074,
  \href{http://dx.doi.org/10.1088/1126-6708/2009/11/074}{\doi{10.1088/1126-6708/2009/11/074}},
\href{http://www.arXiv.org/abs/0908.0631}{\texttt{ arXiv:0908.0631}}.
%%CITATION = ARXIV:0908.0631;%%.

\bibitem{Re:2010bp}
\hrefCMSnoop {} {E.~Re, ``{Single-top Wt-channel production matched with parton
  showers using the POWHEG method}'',} \textit{ Eur. Phys. J.} \textbf{ C71}
  (2011) 1547,
  \href{http://dx.doi.org/10.1140/epjc/s10052-011-1547-z}{\doi{10.1140/epjc/s10052-011-1547-z}},
\href{http://www.arXiv.org/abs/1009.2450}{\texttt{ arXiv:1009.2450}}.
%%CITATION = 1009.2450;%%.

\bibitem{Sjostrand:2006za}
\hrefCMSnoop {} {T.~Sjostrand, S.~Mrenna, and P.~Z. Skands, ``{PYTHIA 6.4
  Physics and Manual}'',} \textit{ JHEP} \textbf{ 05} (2006) 026,
  \href{http://dx.doi.org/10.1088/1126-6708/2006/05/026}{\doi{10.1088/1126-6708/2006/05/026}},
\href{http://www.arXiv.org/abs/hep-ph/0603175}{\texttt{ arXiv:hep-ph/0603175}}.
%%CITATION = HEP-PH/0603175;%%.

\bibitem{Corcella:2000bw}
G.~Corcella\hrefCMSnoop {} { {et~al.}, ``{HERWIG 6: An Event generator for
  hadron emission reactions with interfering gluons (including supersymmetric
  processes)}'',} \textit{ JHEP} \textbf{ 01} (2001) 010,
  \href{http://dx.doi.org/10.1088/1126-6708/2001/01/010}{\doi{10.1088/1126-6708/2001/01/010}},
\href{http://www.arXiv.org/abs/hep-ph/0011363}{\texttt{ arXiv:hep-ph/0011363}}.
%%CITATION = HEP-PH/0011363;%%.

\bibitem{CDF:2014uma}
\hrefCMSnoop {} {{ CDF, D0} Collaboration, ``{Observation of s-channel
  production of single top quarks at the Tevatron}'',} \textit{ Phys. Rev.
  Lett.} \textbf{ 112} (2014) 231803,
  \href{http://dx.doi.org/10.1103/PhysRevLett.112.231803}{\doi{10.1103/PhysRevLett.112.231803}},
\href{http://www.arXiv.org/abs/1402.5126}{\texttt{ arXiv:1402.5126}}.
%%CITATION = ARXIV:1402.5126;%%.

\bibitem{Aad:2015upn}
\hrefCMSnoop {} {{ ATLAS} Collaboration, ``{Evidence for single top-quark
  production in the $s$-channel in proton-proton collisions at $\sqrt{s}=$8 TeV
  with the ATLAS detector using the Matrix Element Method}'',}
\href{http://www.arXiv.org/abs/1511.05980}{\texttt{ arXiv:1511.05980}}.
%%CITATION = ARXIV:1511.05980;%%.

\bibitem{Kondo:1988yd}
\hrefCMSnoop {} {K.~Kondo, ``{Dynamical Likelihood Method for Reconstruction of
  Events With Missing Momentum. 1: Method and Toy Models}'',} \textit{ J. Phys.
  Soc. Jap.} \textbf{ 57} (1988) 4126--4140,
\href{http://dx.doi.org/10.1143/JPSJ.57.4126}{\doi{10.1143/JPSJ.57.4126}}.
%%CITATION = JUPSA,57,4126;%%.

\bibitem{toplhcwg}
\href {https://twiki.cern.ch/twiki/bin/view/LHCPhysics/LHCTopWG}
  {``{https://twiki.cern.ch/twiki/bin/view/LHCPhysics/LHCTopWG}'',}.

\bibitem{Abazov:2011zk}
\hrefCMSnoop {} {{ D0} Collaboration, ``{Precision measurement of the ratio
  $\rm B(t \to Wb)/B(t \to Wq)$ and extraction of $V_{\rm tb}$}'',} \textit{
  Phys. Rev. Lett.} \textbf{ 107} (2011) 121802,
  \href{http://dx.doi.org/10.1103/PhysRevLett.107.121802}{\doi{10.1103/PhysRevLett.107.121802}},
\href{http://www.arXiv.org/abs/1106.5436}{\texttt{ arXiv:1106.5436}}.
%%CITATION = ARXIV:1106.5436;%%.

\bibitem{Aaltonen:2013luz}
\hrefCMSnoop {} {{ CDF} Collaboration, ``{Measurement of $R = \mathcal{B}({\rm
  t \rightarrow Wb})/\mathcal{B}({\rm t \rightarrow Wq})$ in Top--quark--pair
  Decays using Lepton+jets Events and the Full CDF Run II Data set}'',}
  \textit{ Phys. Rev.} \textbf{ D87} (2013), no.~11, 111101,
  \href{http://dx.doi.org/10.1103/PhysRevD.87.111101}{\doi{10.1103/PhysRevD.87.111101}},
\href{http://www.arXiv.org/abs/1303.6142}{\texttt{ arXiv:1303.6142}}.
%%CITATION = ARXIV:1303.6142;%%.

\bibitem{Aaltonen:2014yua}
\hrefCMSnoop {} {{ CDF} Collaboration, ``{Measurement of $B({\rm t \to
  Wb})/B({\rm t \to Wq})$ in Top-Quark-Pair Decays Using Dilepton Events and
  the Full CDF Run II Data Set}'',} \textit{ Phys. Rev. Lett.} \textbf{ 112}
  (2014), no.~22, 221801,
  \href{http://dx.doi.org/10.1103/PhysRevLett.112.221801}{\doi{10.1103/PhysRevLett.112.221801}},
\href{http://www.arXiv.org/abs/1404.3392}{\texttt{ arXiv:1404.3392}}.
%%CITATION = ARXIV:1404.3392;%%.

\bibitem{Khachatryan:2014nda}
\hrefCMSnoop {} {{ CMS} Collaboration, ``{Measurement of the ratio $B({\rm t
  \to Wb})/B({\rm t \to Wq})$ in pp collisions at $\sqrt{s}$ = 8 TeV}'',}
  \textit{ Phys. Lett.} \textbf{ B736} (2014) 33--57,
  \href{http://dx.doi.org/10.1016/j.physletb.2014.06.076}{\doi{10.1016/j.physletb.2014.06.076}},
\href{http://www.arXiv.org/abs/1404.2292}{\texttt{ arXiv:1404.2292}}.
%%CITATION = ARXIV:1404.2292;%%.

\bibitem{Alwall:2006bx}
\hrefCMSnoop {} {J.~Alwall {et~al.}, ``{Is $V_{\rm tb} \approx 1$?}'',}
  \textit{ Eur. Phys. J.} \textbf{ C49} (2007) 791--801,
  \href{http://dx.doi.org/10.1140/epjc/s10052-006-0137-y}{\doi{10.1140/epjc/s10052-006-0137-y}},
\href{http://www.arXiv.org/abs/hep-ph/0607115}{\texttt{ arXiv:hep-ph/0607115}}.
%%CITATION = HEP-PH/0607115;%%.

\bibitem{Lacker:2012ek}
H.~Lacker\hrefCMSnoop {} { {et~al.}, ``{Model-independent extraction of
  $|V_{\rm tq}|$ matrix elements from top-quark measurements at hadron
  colliders}'',} \textit{ Eur. Phys. J.} \textbf{ C72} (2012) 2048,
  \href{http://dx.doi.org/10.1140/epjc/s10052-012-2048-4}{\doi{10.1140/epjc/s10052-012-2048-4}},
\href{http://www.arXiv.org/abs/1202.4694}{\texttt{ arXiv:1202.4694}}.
%%CITATION = ARXIV:1202.4694;%%.

\bibitem{Agashe:2014kda}
\hrefCMSnoop {} {{ Particle Data Group} Collaboration, ``{Review of Particle
  Physics}'',} \textit{ Chin. Phys.} \textbf{ C38} (2014) 090001,
\href{http://dx.doi.org/10.1088/1674-1137/38/9/090001}{\doi{10.1088/1674-1137/38/9/090001}}.
%%CITATION = CHPHD,C38,090001;%%.

\bibitem{Lenz:2013iha}
\hrefCMSnoop {} {A.~Lenz, ``{Constraints on a fourth generation of fermions
  from Higgs Boson searches}'',} \textit{ Adv. High Energy Phys.} \textbf{
  2013} (2013) 910275,
\href{http://dx.doi.org/10.1155/2013/910275}{\doi{10.1155/2013/910275}}.
%%CITATION = 00642,2013,910275;%%.

\bibitem{Okada:2012gy}
\hrefCMSnoop {} {Y.~Okada and L.~Panizzi, ``{LHC signatures of vector-like
  quarks}'',} \textit{ Adv. High Energy Phys.} \textbf{ 2013} (2013) 364936,
  \href{http://dx.doi.org/10.1155/2013/364936}{\doi{10.1155/2013/364936}},
\href{http://www.arXiv.org/abs/1207.5607}{\texttt{ arXiv:1207.5607}}.
%%CITATION = ARXIV:1207.5607;%%.

\bibitem{Aguilar-Saavedra:2013qpa}
\hrefCMSnoop {} {J.~Aguilar-Saavedra, R.~Benbrik, S.~Heinemeyer, and
  M.~Perez-Victoria, ``{Handbook of vector-like quarks: Mixing and single
  production}'',} \textit{ Phys. Rev. D} \textbf{ 88} (2013) 094010,
  \href{http://dx.doi.org/10.1103/PhysRevD.88.094010}{\doi{10.1103/PhysRevD.88.094010}},
\href{http://www.arXiv.org/abs/1306.0572}{\texttt{ arXiv:1306.0572}}.
%%CITATION = ARXIV:1306.0572;%%.

\bibitem{He:1999vp}
\hrefCMSnoop {} {H.-J. He, T.~M.~P. Tait, and C.~P. Yuan, ``{New top flavor
  models with seesaw mechanism}'',} \textit{ Phys. Rev.} \textbf{ D62} (2000)
  011702,
  \href{http://dx.doi.org/10.1103/PhysRevD.62.011702}{\doi{10.1103/PhysRevD.62.011702}},
\href{http://www.arXiv.org/abs/hep-ph/9911266}{\texttt{ arXiv:hep-ph/9911266}}.
%%CITATION = HEP-PH/9911266;%%.

\bibitem{Dobrescu:1997nm}
\hrefCMSnoop {} {B.~A. Dobrescu and C.~T. Hill, ``{Electroweak symmetry
  breaking via top condensation seesaw}'',} \textit{ Phys. Rev. Lett.} \textbf{
  81} (1998) 2634--2637,
  \href{http://dx.doi.org/10.1103/PhysRevLett.81.2634}{\doi{10.1103/PhysRevLett.81.2634}},
\href{http://www.arXiv.org/abs/hep-ph/9712319}{\texttt{ arXiv:hep-ph/9712319}}.
%%CITATION = HEP-PH/9712319;%%.

\bibitem{AguilarSaavedra:2008zc}
\hrefCMSnoop {} {J.~Aguilar-Saavedra, ``{A minimal set of top anomalous
  couplings}'',} \textit{ Nucl.Phys.} \textbf{ B812} (2009) 181--204,
  \href{http://dx.doi.org/10.1016/j.nuclphysb.2008.12.012}{\doi{10.1016/j.nuclphysb.2008.12.012}},
\href{http://www.arXiv.org/abs/0811.3842}{\texttt{ arXiv:0811.3842}}.
%%CITATION = ARXIV:0811.3842;%%.

\bibitem{Georgi:1986ku}
\hrefCMSnoop {} {H.~Georgi, ``{The Flavor Problem}'',} \textit{ Phys. Lett.}
  \textbf{ B169} (1986) 231,
\href{http://dx.doi.org/10.1016/0370-2693(86)90657-X}{\doi{10.1016/0370-2693(86)90657-X}}.
%%CITATION = PHLTA,B169,231;%%.

\bibitem{Zhang:2010dr}
\hrefCMSnoop {} {C.~Zhang and S.~Willenbrock, ``{Effective-Field-Theory
  Approach to Top-Quark Production and Decay}'',} \textit{ Phys. Rev.} \textbf{
  D83} (2011) 034006,
  \href{http://dx.doi.org/10.1103/PhysRevD.83.034006}{\doi{10.1103/PhysRevD.83.034006}},
\href{http://www.arXiv.org/abs/1008.3869}{\texttt{ arXiv:1008.3869}}.
%%CITATION = ARXIV:1008.3869;%%.

\bibitem{Durieux:2014xla}
\hrefCMSnoop {} {G.~Durieux, F.~Maltoni, and C.~Zhang, ``{Global approach to
  top-quark flavor-changing interactions}'',} \textit{ Phys. Rev.} \textbf{
  D91} (2015), no.~7, 074017,
  \href{http://dx.doi.org/10.1103/PhysRevD.91.074017}{\doi{10.1103/PhysRevD.91.074017}},
\href{http://www.arXiv.org/abs/1412.7166}{\texttt{ arXiv:1412.7166}}.
%%CITATION = ARXIV:1412.7166;%%.

\bibitem{Aad:2012gd}
\hrefCMSnoop {} {{ ATLAS} Collaboration, ``{Search for FCNC single top-quark
  production at $\sqrt{s}=7$ TeV with the ATLAS detector}'',} \textit{ Phys.
  Lett.} \textbf{ B712} (2012) 351--369,
  \href{http://dx.doi.org/10.1016/j.physletb.2012.05.022}{\doi{10.1016/j.physletb.2012.05.022}},
\href{http://www.arXiv.org/abs/1203.0529}{\texttt{ arXiv:1203.0529}}.
%%CITATION = ARXIV:1203.0529;%%.

\bibitem{Aad:2015gea}
\hrefCMSnoop {} {{ ATLAS} Collaboration, ``{Search for single top-quark
  production via flavour changing neutral currents at 8 TeV with the ATLAS
  detector}'',} (2015).
\href{http://www.arXiv.org/abs/1509.00294}{\texttt{ arXiv:1509.00294}}.
%%CITATION = ARXIV:1509.00294;%%.

\bibitem{Khachatryan:2015att}
\hrefCMSnoop {} {{ CMS} Collaboration, ``{Search for anomalous single top quark
  production in association with a photon in pp collisions at $\sqrt{s} = 8$
  TeV}'',}
\href{http://www.arXiv.org/abs/1511.03951}{\texttt{ arXiv:1511.03951}}.
%%CITATION = ARXIV:1511.03951;%%.

\bibitem{Achard:2002vv}
\hrefCMSnoop {} {{ L3} Collaboration, ``{Search for single top production at
  LEP}'',} \textit{ Phys. Lett.} \textbf{ B549} (2002) 290--300,
  \href{http://dx.doi.org/10.1016/S0370-2693(02)02933-7}{\doi{10.1016/S0370-2693(02)02933-7}},
\href{http://www.arXiv.org/abs/hep-ex/0210041}{\texttt{ arXiv:hep-ex/0210041}}.
%%CITATION = HEP-EX/0210041;%%.

\bibitem{Abramowicz:2011tv}
\hrefCMSnoop {} {{ ZEUS} Collaboration, ``{Search for single-top production in
  ep collisions at HERA}'',} \textit{ Phys. Lett.} \textbf{ B708} (2012)
  27--36,
  \href{http://dx.doi.org/10.1016/j.physletb.2012.01.025}{\doi{10.1016/j.physletb.2012.01.025}},
\href{http://www.arXiv.org/abs/1111.3901}{\texttt{ arXiv:1111.3901}}.
%%CITATION = ARXIV:1111.3901;%%.

\bibitem{Aaron:2009vv}
\hrefCMSnoop {} {{ H1} Collaboration, ``{Search for Single Top Quark Production
  at HERA}'',} \textit{ Phys. Lett.} \textbf{ B678} (2009) 450--458,
  \href{http://dx.doi.org/10.1016/j.physletb.2009.06.057}{\doi{10.1016/j.physletb.2009.06.057}},
\href{http://www.arXiv.org/abs/0904.3876}{\texttt{ arXiv:0904.3876}}.
%%CITATION = ARXIV:0904.3876;%%.

\bibitem{Abazov:2011qf}
\hrefCMSnoop {} {{ D0} Collaboration, ``{Search for flavor changing neutral
  currents in decays of top quarks}'',} \textit{ Phys. Lett.} \textbf{ B701}
  (2011) 313--320,
  \href{http://dx.doi.org/10.1016/j.physletb.2011.06.014}{\doi{10.1016/j.physletb.2011.06.014}},
\href{http://www.arXiv.org/abs/1103.4574}{\texttt{ arXiv:1103.4574}}.
%%CITATION = ARXIV:1103.4574;%%.

\bibitem{Aaltonen:2008ac}
\hrefCMSnoop {} {{ CDF} Collaboration, ``{Search for the Flavor Changing
  Neutral Current Decay $\rm t \to Zq$ in $\rm p\bar{p}$ Collisions at
  $\sqrt{s} = 1.96$ TeV}'',} \textit{ Phys. Rev. Lett.} \textbf{ 101} (2008)
  192002,
  \href{http://dx.doi.org/10.1103/PhysRevLett.101.192002}{\doi{10.1103/PhysRevLett.101.192002}},
\href{http://www.arXiv.org/abs/0805.2109}{\texttt{ arXiv:0805.2109}}.
%%CITATION = ARXIV:0805.2109;%%.

\bibitem{Aad:2012ij}
\hrefCMSnoop {} {{ ATLAS} Collaboration, ``{A search for flavour changing
  neutral currents in top-quark decays in pp collision data collected with the
  ATLAS detector at $\sqrt{s}=7$~TeV}'',} \textit{ JHEP} \textbf{ 09} (2012)
  139,
  \href{http://dx.doi.org/10.1007/JHEP09(2012)139}{\doi{10.1007/JHEP09(2012)139}},
\href{http://www.arXiv.org/abs/1206.0257}{\texttt{ arXiv:1206.0257}}.
%%CITATION = ARXIV:1206.0257;%%.

\bibitem{Chatrchyan:2013nwa}
\hrefCMSnoop {} {{ CMS} Collaboration, ``{Search for Flavor-Changing Neutral
  Currents in Top-Quark Decays $\rm t \to Zq$ in pp Collisions at
  $\sqrt{s}=8$~ TeV}'',} \textit{ Phys. Rev. Lett.} \textbf{ 112} (2014),
  no.~17, 171802,
  \href{http://dx.doi.org/10.1103/PhysRevLett.112.171802}{\doi{10.1103/PhysRevLett.112.171802}},
\href{http://www.arXiv.org/abs/1312.4194}{\texttt{ arXiv:1312.4194}}.
%%CITATION = ARXIV:1312.4194;%%.

\bibitem{PhysRevLett.80.2525}
\hrefCMSnoop {} {{ CDF} Collaboration, ``{Search for flavor-changing neutral
  current decays of the top quark in $\rm p \bar{p}$ collisions at $\sqrt{s} =
  1.8$ TeV}'',} \textit{ Phys. Rev. Lett.} \textbf{ 80} (1998) 2525--2530,
\href{http://dx.doi.org/10.1103/PhysRevLett.80.2525}{\doi{10.1103/PhysRevLett.80.2525}}.
%%CITATION = PRLTA,80,2525;%%.

\bibitem{Mahlon:1999gz}
\hrefCMSnoop {} {G.~Mahlon and S.~J. Parke, ``{Single top quark production at
  the LHC: Understanding spin}'',} \textit{ Phys.Lett.} \textbf{ B476} (2000)
  323--330,
  \href{http://dx.doi.org/10.1016/S0370-2693(00)00149-0}{\doi{10.1016/S0370-2693(00)00149-0}},
\href{http://www.arXiv.org/abs/hep-ph/9912458}{\texttt{ arXiv:hep-ph/9912458}}.
%%CITATION = HEP-PH/9912458;%%.

\bibitem{Jezabek:1994zv}
\hrefCMSnoop {} {M.~Jezabek and J.~H. Kuhn, ``{V--A tests through leptons from
  polarised top quarks}'',} \textit{ Phys.Lett.} \textbf{ B329} (1994)
  317--324,
  \href{http://dx.doi.org/10.1016/0370-2693(94)90779-X}{\doi{10.1016/0370-2693(94)90779-X}},
\href{http://www.arXiv.org/abs/hep-ph/9403366}{\texttt{ arXiv:hep-ph/9403366}}.
%%CITATION = HEP-PH/9403366;%%.

\bibitem{AguilarSaavedra:2010nx}
\hrefCMSnoop {} {J.~Aguilar-Saavedra and J.~Bernabeu, ``{W polarisation beyond
  helicity fractions in top quark decays}'',} \textit{ Nucl.Phys.} \textbf{
  B840} (2010) 349--378,
  \href{http://dx.doi.org/10.1016/j.nuclphysb.2010.07.012}{\doi{10.1016/j.nuclphysb.2010.07.012}},
\href{http://www.arXiv.org/abs/1005.5382}{\texttt{ arXiv:1005.5382}}.
%%CITATION = ARXIV:1005.5382;%%.

\bibitem{AguilarSaavedra:2008gt}
\hrefCMSnoop {} {J.~Aguilar-Saavedra, ``{Single top quark production at LHC
  with anomalous Wtb couplings}'',} \textit{ Nucl.Phys.} \textbf{ B804} (2008)
  160--192,
  \href{http://dx.doi.org/10.1016/j.nuclphysb.2008.06.013}{\doi{10.1016/j.nuclphysb.2008.06.013}},
\href{http://www.arXiv.org/abs/0803.3810}{\texttt{ arXiv:0803.3810}}.
%%CITATION = ARXIV:0803.3810;%%.

\bibitem{Bach:2012fb}
\hrefCMSnoop {} {F.~Bach and T.~Ohl, ``{Anomalous top couplings at hadron
  colliders revisited}'',} \textit{ Phys.Rev.} \textbf{ D86} (2012) 114026,
  \href{http://dx.doi.org/10.1103/PhysRevD.86.114026}{\doi{10.1103/PhysRevD.86.114026}},
\href{http://www.arXiv.org/abs/1209.4564}{\texttt{ arXiv:1209.4564}}.
%%CITATION = ARXIV:1209.4564;%%.

\bibitem{CMS-PAS-TOP-13-001}
\hrefCMSnoop {} {{ CMS} Collaboration, ``{Measurement of top quark polarisation
  in t-channel single top quark production}'',}
\href{http://www.arXiv.org/abs/1511.02138}{\texttt{ arXiv:1511.02138}}.
%%CITATION = ARXIV:1511.02138;%%.

\bibitem{Khachatryan:2014vma}
\hrefCMSnoop {} {{ CMS} Collaboration, ``{Measurement of the W boson helicity
  in events with a single reconstructed top quark in pp collisions at $
  \sqrt{s}=8 $ TeV}'',} \textit{ JHEP} \textbf{ 01} (2015) 053,
  \href{http://dx.doi.org/10.1007/JHEP01(2015)053}{\doi{10.1007/JHEP01(2015)053}},
\href{http://www.arXiv.org/abs/1410.1154}{\texttt{ arXiv:1410.1154}}.
%%CITATION = ARXIV:1410.1154;%%.

\bibitem{CMS-PAS-TOP-12-025}
\href {https://cds.cern.ch/record/1528567} {{ATLAS and CMS Collaborations},
  ``{LHC Combination note: W helicities}'',} Technical Report
  CMS-PAS-TOP-12-025 and ATLAS-CONF-2013-033, CERN, Geneva, 2013.

\bibitem{Aad:2015yem}
\hrefCMSnoop {} {{ ATLAS} Collaboration, ``{Search for anomalous couplings in
  the Wtb vertex from the measurement of double differential angular decay
  rates of single top quarks produced in the t-channel with the ATLAS
  detector}'',}
\href{http://www.arXiv.org/abs/1510.03764}{\texttt{ arXiv:1510.03764}}.
%%CITATION = ARXIV:1510.03764;%%.

\bibitem{AguilarSaavedra:2012xe}
\hrefCMSnoop {} {J.~A. Aguilar-Saavedra and R.~V. Herrero-Hahn,
  ``{Model-independent measurement of the top quark polarisation}'',} \textit{
  Phys. Lett.} \textbf{ B718} (2013) 983--987,
  \href{http://dx.doi.org/10.1016/j.physletb.2012.11.031}{\doi{10.1016/j.physletb.2012.11.031}},
\href{http://www.arXiv.org/abs/1208.6006}{\texttt{ arXiv:1208.6006}}.
%%CITATION = ARXIV:1208.6006;%%.

\bibitem{BORDES1993315}
\hrefCMSnoop {} {G.~Bordes and B.~van Eijk, ``{On the associate production of a
  neutral intermediate-mass Higgs boson with a single top quark at the LHC and
  SSC}'',} \textit{ Physics Letters B} \textbf{ 299} (1993), no.~3, 315 -- 320,
  \href{http://dx.doi.org/http://dx.doi.org/10.1016/0370-2693(93)90266-K}{\doi{http://dx.doi.org/10.1016/0370-2693(93)90266-K}}.

\bibitem{Hespel:2015zea}
\hrefCMSnoop {} {B.~Hespel, F.~Maltoni, and E.~Vryonidou, ``{Higgs and Z boson
  associated production via gluon fusion in the SM and the 2HDM}'',} \textit{
  JHEP} \textbf{ 06} (2015) 065,
  \href{http://dx.doi.org/10.1007/JHEP06(2015)065}{\doi{10.1007/JHEP06(2015)065}},
\href{http://www.arXiv.org/abs/1503.01656}{\texttt{ arXiv:1503.01656}}.
%%CITATION = ARXIV:1503.01656;%%.

\bibitem{Farina:2012xp}
M.~Farina\hrefCMSnoop {} { {et~al.}, ``{Lifting degeneracies in Higgs couplings
  using single top production in association with a Higgs boson}'',} \textit{
  JHEP} \textbf{ 05} (2013) 022,
  \href{http://dx.doi.org/10.1007/JHEP05(2013)022}{\doi{10.1007/JHEP05(2013)022}},
\href{http://www.arXiv.org/abs/1211.3736}{\texttt{ arXiv:1211.3736}}.
%%CITATION = ARXIV:1211.3736;%%.

\bibitem{Maltoni2001}
\hrefCMSnoop {} {F.~Maltoni, K.~Paul, and S.~Willenbrock, ``Associated
  production of Higgs and single top at hadron colliders'',} \textit{ Phys.
  Rev. D} \textbf{ 64} (2001) 094023,
  \href{http://dx.doi.org/10.1103/PhysRevD.64.094023}{\doi{10.1103/PhysRevD.64.094023}},
\href{http://www.arXiv.org/abs/hep-ph/0106293}{\texttt{ arXiv:hep-ph/0106293}}.
%%CITATION = HEP-PH/0106293;%%.

\bibitem{Heinemeyer:2013tqa}
\hrefCMSnoop {} {{ LHC Higgs Cross Section Working Group} Collaboration,
  ``{Handbook of LHC Higgs Cross Sections: 3. Higgs Properties}'',}
  \href{http://dx.doi.org/10.5170/CERN-2013-004}{\doi{10.5170/CERN-2013-004}},
\href{http://www.arXiv.org/abs/1307.1347}{\texttt{ arXiv:1307.1347}}.
%%CITATION = ARXIV:1307.1347;%%.

\bibitem{Biswas:2012bd}
\hrefCMSnoop {} {S.~Biswas, E.~Gabrielli, and B.~Mele, ``{Single top and Higgs
  associated production as a probe of the Htt coupling sign at the LHC}'',}
  \textit{ JHEP} \textbf{ 01} (2013) 088,
  \href{http://dx.doi.org/10.1007/JHEP01(2013)088}{\doi{10.1007/JHEP01(2013)088}},
\href{http://www.arXiv.org/abs/1211.0499}{\texttt{ arXiv:1211.0499}}.
%%CITATION = ARXIV:1211.0499;%%.

\bibitem{Biswas:2013xva}
\hrefCMSnoop {} {S.~Biswas, E.~Gabrielli, F.~Margaroli, and B.~Mele, ``{Direct
  constraints on the top-Higgs coupling from the 8 TeV LHC data}'',} \textit{
  JHEP} \textbf{ 07} (2013) 073,
  \href{http://dx.doi.org/10.1007/JHEP07(2013)073}{\doi{10.1007/JHEP07(2013)073}},
\href{http://www.arXiv.org/abs/1304.1822}{\texttt{ arXiv:1304.1822}}.
%%CITATION = ARXIV:1304.1822;%%.

\bibitem{Chang:2014rfa}
\hrefCMSnoop {} {J.~Chang, K.~Cheung, J.~S. Lee, and C.-T. Lu, ``{Probing the
  Top-Yukawa Coupling in Associated Higgs production with a Single Top
  Quark}'',} \textit{ JHEP} \textbf{ 05} (2014) 062,
  \href{http://dx.doi.org/10.1007/JHEP05(2014)062}{\doi{10.1007/JHEP05(2014)062}},
\href{http://www.arXiv.org/abs/1403.2053}{\texttt{ arXiv:1403.2053}}.
%%CITATION = ARXIV:1403.2053;%%.

\bibitem{CMS-PAS-HIG-14-027}
\hrefCMSnoop {} {{ CMS} Collaboration, ``{Search for the associated production
  of a Higgs boson with a single top quark in proton-proton collisions at
  $\sqrt{s} = 8$~TeV}'',} (2015).
\href{http://www.arXiv.org/abs/1509.08159}{\texttt{ arXiv:1509.08159}}.
%%CITATION = ARXIV:1509.08159;%%.

\bibitem{Aad:2014lma}
\hrefCMSnoop {} {{ ATLAS} Collaboration, ``{Search for $\rm H \to \gamma\gamma$
  produced in association with top quarks and constraints on the Yukawa
  coupling between the top quark and the Higgs boson using data taken at 7 TeV
  and 8 TeV with the ATLAS detector}'',} \textit{ Phys. Lett.} \textbf{ B740}
  (2015) 222--242,
  \href{http://dx.doi.org/10.1016/j.physletb.2014.11.049}{\doi{10.1016/j.physletb.2014.11.049}},
\href{http://www.arXiv.org/abs/1409.3122}{\texttt{ arXiv:1409.3122}}.
%%CITATION = ARXIV:1409.3122;%%.

\bibitem{Kidonakis:2013zqa}
\hrefCMSnoop {} {N.~Kidonakis, ``{Top Quark Production}'',} in \textit{
  {Proceedings, Helmholtz International Summer School on Physics of Heavy
  Quarks and Hadrons (HQ 2013)}}, pp.~139--168.
\newblock 2014.
\newblock \href{http://www.arXiv.org/abs/1311.0283}{\texttt{ arXiv:1311.0283}}.
\newblock
\href{http://dx.doi.org/10.3204/DESY-PROC-2013-03/Kidonakis}{\doi{10.3204/DESY-PROC-2013-03/Kidonakis}}.
%%CITATION = ARXIV:1311.0283;%%.

\bibitem{fourfiveflavorschemesmalt}
\hrefCMSnoop {} {F.~Maltoni, G.~Ridolfi, and M.~Ubiali, ``{b-initiated
  processes at the LHC: a reappraisal}'',} \textit{ JHEP} \textbf{ 04} (2012)
  022,
  \href{http://dx.doi.org/10.1007/JHEP07(2012)022}{\doi{10.1007/JHEP07(2012)022}},
  \href{http://www.arXiv.org/abs/1203.6393}{\texttt{ arXiv:1203.6393}}.

\bibitem{fourfiveflavorschemescamp}
\hrefCMSnoop {} {J.~Campbell, R.~Frederix, F.~Maltoni, and F.~Tramontano,
  ``{NLO predictions for t-channel production of single top and fourth
  generation quarks at hadron colliders}'',} \textit{ JHEP} \textbf{ 0910}
  (2009) 042,
  \href{http://dx.doi.org/10.1088/1126-6708/2009/10/042}{\doi{10.1088/1126-6708/2009/10/042}},
  \href{http://www.arXiv.org/abs/0907.3933}{\texttt{ arXiv:0907.3933}}.

\bibitem{Czakon:2011xx}
\hrefCMSnoop {} {M.~Czakon and A.~Mitov, ``{Top++: A Program for the
  Calculation of the Top-Pair Cross-Section at Hadron Colliders}'',} \textit{
  Comput. Phys. Commun.} \textbf{ 185} (2014) 2930,
  \href{http://dx.doi.org/10.1016/j.cpc.2014.06.021}{\doi{10.1016/j.cpc.2014.06.021}},
\href{http://www.arXiv.org/abs/1112.5675}{\texttt{ arXiv:1112.5675}}.
%%CITATION = ARXIV:1112.5675;%%.

\bibitem{Campbell:2013yla}
\hrefCMSnoop {} {J.~Campbell, R.~K. Ellis, and R.~Rontsch, ``{Single top
  production in association with a Z boson at the LHC}'',} \textit{ Phys.Rev.
  D} \textbf{ 87} (2013) 114006,
  \href{http://dx.doi.org/10.1103/PhysRevD.87.114006}{\doi{10.1103/PhysRevD.87.114006}},
\href{http://www.arXiv.org/abs/1302.3856}{\texttt{ arXiv:1302.3856}}.
%%CITATION = ARXIV:1302.3856;%%.

\bibitem{Demartin:2015uha}
\hrefCMSnoop {} {F.~Demartin, F.~Maltoni, K.~Mawatari, and M.~Zaro, ``{Higgs
  production in association with a single top quark at the LHC}'',} \textit{
  Eur.Phys.J. C} \textbf{ 75} (2015) 267,
  \href{http://dx.doi.org/10.1140/epjc/s10052-015-3475-9}{\doi{10.1140/epjc/s10052-015-3475-9}},
\href{http://www.arXiv.org/abs/1504.00611}{\texttt{ arXiv:1504.00611}}.
%%CITATION = ARXIV:1504.00611;%%.

\end{thebibliography}\endgroup

\end{document}